\documentclass[twocolumn,showpacs,nofootinbib,preprintnumbers,amsmath,amssymb]{revtex4-1}

\usepackage{amsfonts}
\usepackage{amsmath}
\usepackage{amssymb}
\usepackage{amstext}
\usepackage{amsthm}
\usepackage{graphicx}
\usepackage{dcolumn}
\usepackage{bm}
\usepackage{epsf}
\usepackage{framed}
\usepackage{hyperref}

\begin{document}


\title{Unbiased sampling of network ensembles}

\author{Tiziano Squartini}
\affiliation{Instituut-Lorentz for Theoretical Physics, University of Leiden, Niels Bohrweg 2, 2333 CA Leiden (The Netherlands)\\
Institute for Complex Systems UOS Sapienza,
``Sapienza" University of Rome, P.le Aldo Moro 5, 00185 Rome (Italy)}
\author{Rossana Mastrandrea}
\affiliation{Institute of Economics and LEM, Scuola Superiore Sant'Anna, 56127 Pisa (Italy)\\
Aix Marseille Universit\'e, Universit\'e de Toulon, CNRS, CPT, UMR 7332, 13288 Marseille (France)}
\author{Diego Garlaschelli}
\affiliation{Instituut-Lorentz for Theoretical Physics, University of Leiden, Niels Bohrweg 2, 2333 CA Leiden (The Netherlands)}
%


\date{\today}

\begin{abstract}
Sampling random graphs with given properties is a key step in the analysis of networks, as random ensembles represent basic null models required to identify patterns such as communities and motifs. An important requirement is that the sampling process is unbiased and efficient. The main approaches are microcanonical, i.e. they sample graphs that match the enforced constraints exactly. Unfortunately, when applied to strongly heterogeneous networks (like most real-world examples), the majority of these approaches become biased and/or time-consuming. 
Moreover, the algorithms defined in the simplest cases, such as binary graphs with given degrees, are not easily generalizable to more complicated ensembles. Here we propose a solution to the problem via the introduction of a ``Maximize and Sample'' (``Max \& Sam'' for short) method to correctly sample ensembles of networks where the constraints are `soft', i.e.   realized as ensemble averages. Our method is based on exact maximum-entropy distributions and is therefore unbiased by construction, even for strongly heterogeneous networks. It is also more computationally efficient than most microcanonical alternatives. Finally, it works for both binary and weighted networks with a variety of constraints, including combined degree-strength sequences and full reciprocity structure, for which no alternative method exists.
Our canonical approach can in principle be turned into an unbiased microcanonical one, via a restriction to the relevant subset. Importantly, the analysis of the fluctuations of the constraints suggests that the microcanonical and canonical versions of all the ensembles considered here are not equivalent. 
We show various real-world applications and provide a code implementing all our algorithms.
\end{abstract}

\maketitle
\section{Introduction}
Network theory is systematically used to address problems of scientific and societal relevance \cite{Newman 2010}, from the prediction of the spreading of infectious diseases worldwide \cite{Colizza et al 2006} to the identification of early-warning signals of upcoming financial crises \cite{Squartini et al 2013}. 
More in general, several dynamical and stochastic processes are strongly affected by the topology of the underlying network \cite{Barrat 2008}. 
This results in the need to identify the topological properties that are statistically significant in a real network, i.e. to discriminate which higher-order properties can be directly traced back to the local features of nodes, and which are instead due to additional factors.

To achieve this goal, one requires (a family of) randomized benchmarks, i.e. ensembles of graphs where the local heterogeneity is the same as in the real network, and the topology is random in any other respect: this defines a \emph{null model} of the original network. 
Nontrivial patterns can then be detected in the form of empirical deviations from the theoretical expectations of the null model \cite{mymethod}. 
Important examples of such patterns is the presence of \emph{motifs} (recurring subgraphs of small size, like building blocks of a network \cite{motifs}) and \emph{communities} (groups of nodes that are more densely connected internally than with each other \cite{communities}).
To detect these and many other patterns, one needs to correctly specify the null model and then calculate e.g. the average and standard deviation (or alternatively a
confidence interval) of any topological property of interest over the corresponding randomized ensemble of graphs. 

Unfortunately, given the strong heterogeneity of nodes (e.g. the power-law distribution of vertex degrees), the solution to the above problem is not simple. 
This is most easily explained in the case of binary graphs, even if similar arguments apply to weighted networks as well. 
For simple graphs, the most important null model is the (Undirected Binary) Configuration Model (UBCM), defined as an ensemble of networks where the degree of each node is specified, and the rest of the topology is maximally random \cite{bollobas,molloy,Newman et al 2001}. 
Since the degrees of all nodes (the so-called \emph{degree sequence}) act as constraints, ``maximally random'' does not mean ``completely random'': in order to realize the degree
sequence, interdependencies among vertices necessarily arise. These interdependencies affect other topological properties as well. So, even if the degree sequence is the only quantity that is enforced `on purpose',
other structural properties are unavoidably constrained as well. These higher-order effects are called
``structural correlations''. In order to disentangle spurious structural correlations from genuine correlations of interest, 
it is very important to properly implement the UBCM in such a way that it takes the observed degree sequence as input and generates expectations based on a uniform and efficient sampling of the ensemble.
Similar and more challenging considerations apply to other null models, defined e.g. for directed or weighted graphs and specified by more general constraints.

Several approaches to the problem have been proposed and can be roughly divided in two large classes: \emph{microcanonical} and \emph{canonical} methods.
Microcanonical approaches \cite{MS,coolen2,coolen,Artzy-Randrup et al 2005,Del Genio 2010,Kim et al 2012,Blitzstein 2011} aim at artificially generating many randomized variants of the observed network in such a way that the constrained properties are identical to the empirical ones, thus creating a collection of graphs sampling the desired ensemble. In these algorithms the enforced constraints are `hard', i.e they are met exactly by each graph in the resulting ensemble. 
As we discuss in this paper, this strong requirement implies that most microcanonical approaches proposed so far suffer from various problems, including bias, lack of ergodicity, mathematical intractability, high computational demands, and poor generalizability.

On the other hand, in \emph{canonical} approaches \cite{mymethod,newman_expo,ginestra,fronczak,myPRE1,myPRE2,mynull,mygrandcanonical,mymotifs,mywreciprocity,mybosefermi,myenhanced} the constraints are `soft', i.e. they can be violated by individual graphs in the ensemble, even if the ensemble average of each constraint still matches the enforced value exactly. 
Canonical approaches are generally introduced to directly obtain, as a function of the observed constraints (e.g. the degree sequence), exact mathematical expressions for the expected topological properties, thus avoiding the explicit generation of randomized networks \cite{mymethod}. 
However, this is only possible if the mathematical expressions for the topological properties of interest are simple enough to make the analytical calculation of the expected values feasible. 
Unfortunately, the most popular approaches rely on highly approximated expressions leading to ill-defined or unknown probabilities that cannot be used to sample the ensemble. These approximations are in any case available only for the simplest ensembles (e.g. the UBCM), leaving the problem unsolved for more general constraints. 
This implies that the computational use of canonical null models has not been implemented systematically so far.

In this paper, by combining an exact maximum-likelihood approach with an efficient computational sampling scheme, we define a rigorously unbiased method to sample  ensembles of various types of networks (i.e. directed, undirected, weighted, binary) with many possible constraints (degree sequence, strength sequence, reciprocity structure, mixed binary and weighted properties, etc.).
We make use of a series of recent analytical results that generate the exact probabilities in all these cases  of interest \cite{mymethod,myPRE1,myPRE2,mynull,mygrandcanonical,mymotifs,mywreciprocity,mybosefermi,myenhanced,myextensive}
and consider various examples illustrating the usefulness of our method when applied to real-world networks.
 
We also analyse the canonical fluctuations of the constraints in each model.
Previous theoretical analyses of fluctuations in some network ensembles have been carried out, for instance, in ref.\cite{gin1} for graphs with given degree sequence and in ref. \cite{gin2} for graphs with given community structure. 
Also, a comparison between some microcanonical and canonical network ensembles has been carried out in ref. \cite{gin3}. 
In this paper, we provide a complete analytical characterization of the fluctuations of each constraint for all the ensembles under study. For the majority of these ensembles, the exact analytical expressions characterizing the fluctuations are derived here for the first time.
Moreover, in our maximum-likelihood approach the knowledge of the hidden variables allows us to calculate, for the first time, the exact value of the fluctuations explicitly for each node in the empirical networks considered.
Our results suggest that, unlike in most physical systems, the microcanonical and canonical versions of the graph ensembles considered here are surprisingly  \emph{not} equivalent (see ref. \cite{mybreaking} for a recent mathematical proof of ensemble nonequivalence in the UBCM).

In any case, our canonical method can in principle be converted into an unbiased microcanonical one, if we discard all the sampled networks that violate the sharp constraints. At the end of the paper, we discuss the advantages and disadvantages of this procedure explicitly, and clarify that canonical ensembles are more appropriate in presence of missing entries or errors in the data.

Finally, we include an appendix with a description of a algorithm that we have explicitly coded in various ways \cite{routine_address,routine_address2,routine_address3}.
The algorithm allows the users to sample all the graph ensembles described in this paper, given an empirically observed network (or even only the values of the constraints).

\section{Previous approaches\label{sec:previous}}
In this section, we briefly discuss the main available approaches to the problem of sampling network ensembles with given constraints, and highlight the limitations that call for an improved solution. 
We consider both microcanonical and canonical methods.
In both cases, since the UBCM is the most popular and most studied ensemble, we will discuss the problem by focusing mainly on the implementations of this model. The same kind of considerations extend to other constraints  and other types of networks as well.

\subsection{Microcanonical methods\label{sec:micro}}
There have been several attempts to develop microcanonical algorithms that efficiently implement the UBCM. 
One of the earliest algorithm starts with an empty network having the same number of vertices of
the original one, where each vertex is assigned a number of `half edges' (or `edge stubs') equal to its degree in the real network. Then, pairs of stubs are randomly matched, thus creating
the final edges of a random network with the desired degree sequence \cite{Newman et al 2001}. 
Unfortunately, for most empirical networks, the heterogeneity of the degrees is such that this algorithm produces several
multiple edges between vertices with large degree, and several self-loops \cite{MS}. 
If the formation of these undesired edges is forbidden explicitly, the algorithm gets stuck in configurations where
edge stubs have no more eligible partners, thus failing to complete any randomized network. 

To overcome this limitation, a different algorithm (which is still widely used) was introduced \cite{MS}.
This ``Local Rewiring Algorithm'' (LRA) starts from the original network, rather than from scratch, and
randomizes the topology through the iteration of an elementary move that preserves the degrees of all nodes. While this algorithm always produces random networks, it is very time consuming since
many iterations of the fundamental move are needed in order to produce just one randomized variant, and
this entire operation has to be repeated several times (the mixing time being still unknown \cite{Milo et al 2003}) in order to produce
many variants. 

Besides these practical problems, the main conceptual limitation of the LRA is the fact that it is \emph{biased},
i.e. it does not sample the desired ensemble uniformly. This has been rigorously shown relatively recently
\cite{Artzy-Randrup et al 2005,coolen2,coolen}. For undirected networks, uniformity has been shown to hold, at
least approximately, only when the degree sequence is such that \cite{coolen2}
\begin{equation}
k_{max}\cdot{\overline{k^2}}/{(\overline{k})^2}\ll N
\label{eq_1}
\end{equation}
\noindent where $k_{max}$ is the largest degree in the network, $\overline{k}$ is the average degree, $\overline{k^2}$ is the second moment, and $N$ is the number of vertices. 
Clearly, the above condition sets an upper bound for the heterogeneity of the degrees of vertices, and is violated if the heterogeneity is strong. This is a first indication that
the available methods break down for `strongly heterogeneous' networks. 
As we discuss later, most real-world networks are known to fall precisely within this class.
For directed networks, where links are oriented and the constraints to be met are the numbers of incoming and outgoing links (in-degree and out-degree) separately, a condition similar to eq.(\ref{eq_1}) is required to avoid the generation of bias \cite{coolen}. 
Again, this condition is strongly violated by most real-world networks.
Moreover, the directed version of the LRA is also non-ergodic, i.e. it is in general not able to explore the entire ensemble of networks \cite{coolen}. 

It has been shown that ergodicity can be restored by introducing an additional triangular move inverting the direction of closed loops of three vertices \cite{coolen}.
However, in order to restore uniformity (for both directed and undirected graphs) one needs to introduce an appropriate acceptance probability for the rewiring move \cite{Artzy-Randrup
et al 2005,coolen2,coolen}.
Unfortunately, the acceptance probability depends on some nontrivial property of the current network configuration. Since this property must be recalculated at each step, the
resulting algorithm is significantly time consuming.
Quantifying the bias generated by the LRA when eq.(\ref{eq_1}) (or its directed counterpart) is violated is difficult, mainly because an exact mathematical characterization of microcanonical graph ensembles valid in such regime is still lacking.
Yet, the proof of the existence of bias provided in refs. \cite{coolen2,coolen} is an obvious warning against the use of the LRA on strongly heterogeneous networks. The reader is referred to those papers for a discussion.

Other recent alternatives \cite{Del Genio 2010, Kim et al 2012, Blitzstein 2011} rely on theorems, such as the Erd\H{o}s-Gallai \cite{EG} one, that set necessary and sufficient conditions for a degree sequence to be \emph{graphic}, i.e. realized by at least one graph. 
These `graphic' methods exploit such (or related) conditions to define biased sampling algorithms in conjunction with the estimation of the corresponding sampling probabilities, thus allowing one to statistically reweight the outcome and sample the ensemble effectively uniformly \cite{Del Genio 2010, Kim et al 2012, Blitzstein 2011}.
Del Genio et al. \cite{Del Genio 2010} show that, for networks with power-law degree distribution of the form $P(k)\sim k^{-\gamma}$, the computational complexity of sampling \emph{just one} graph using their algorithm is $O(N^2)$ if $\gamma>3$.
However, when $\gamma<3$ the computational complexity increases to $O(N^{2.5})$ if 
\begin{equation}
k_{max}<\sqrt{N}
\label{eq_0}
\end{equation}
and to $O(N^{3})$ if $k_{max}>\sqrt{N}$.
The upper bound $\sqrt{N}$ is a particular case of the so-called ``structural cut-off'' that we will discuss in more detail later.
For the moment, it is enough for us to note that eq.\eqref{eq_0} is another indication that, for strongly heterogeneous networks, the problem of sampling becomes more complicated.
Unfortunately, most real networks violate eq.\eqref{eq_0} strongly.

So, while `graphic' algorithms do provide a solution for every network, their complexity increases for networks of increasing (and more realistic) heterogeneity. A more fundamental limitation is that these methods can only handle the problem of binary graphs with given degree sequence. 
The generalization to other types of networks and other constraints is not straightforward, as it would require the proof of more general `graphicality' theorems, and \emph{ad hoc} modifications of the algorithm.

\subsection{Canonical methods\label{sec:macro}}
Canonical approaches aim at obtaining, as a function of the observed constraints (e.g. the degree sequence), mathematical expressions for the expected topological properties, avoiding the explicit generation of randomized networks. 
For canonical methods the requirement of uniformity is replaced by the requirement that the proability distribution over the enlarged ensemble has maximum entropy \cite{mymethod,newman_expo}.

For binary graphs, since any topological property $X$  is a function $X(\mathbf{A})$ of the adjacency matrix $\mathbf{A}$ of the network (with entries $a_{ij} = 1$ if the vertices $i$ and $j$ are connected, and $a_{ij} = 0$ otherwise), the ultimate goal is that of finding a mathematical expression for the probability $P(\mathbf{A})$ of occurrence of each graph.
This allows to compute the expected value of $X$ as $\sum_\mathbf{A}P(\mathbf{A})X(\mathbf{A})$.
 Importantly, for canonical ensembles with local constraints $P(\mathbf{A})$ factorizes to a product over pairs of nodes, where each term in the product involves the probability $p_{ij}$ that the vertices $i$ and $j$ are connected in the ensemble. 
Determining the mathematical form of $p_{ij}$ is the main goal of canonical approaches.
Note that, by contrast, in the microcanonical ensemble all links are dependent on each other (the degree sequence must be reproduced exactly in each realization), which implies that the probability of the entire graph does not factorize to node-pair probabilities. 

For binary undirected networks, the most popular specification for $p_{ij}$ is the  factorized one \cite{Newman 2010,Chung et al 2002a,Chung et al 2002b}:
\begin{equation}
p_{ij}=\frac{k_i k_j}{k_{tot}}
\label{eq_2}
\end{equation}
(where $k_i$ is the degree of node $i$ and $k_{tot}$ is the total degree over all nodes).
For weighted undirected networks, where each link can have a non-negative weight $w_{ij}$ and each
vertex $i$ is characterized by a given strength $s_i$ (the total weight of the links of node $i$), the corresponding assumption is that the expected weight of the link connecting the vertices $i$ and $j$ is
\begin{equation}
\langle w_{ij}\rangle=\frac{s_i s_j}{s_{tot}}
\label{eq_3}
\end{equation}
(where $s_{tot}$ is the total strength of all vertices). 

Equations \eqref{eq_2} and \eqref{eq_3} are routinely used, and have become
standard textbook expressions \cite{Newman 2010}. The most frequent use of these expressions is perhaps encountered
in the empirical analysis of \emph{communities}, i.e. relatively denser modules of vertices
in large networks \cite{communities}. 
Most community detection algorithms compare different partitions of vertices into communities (each partition being parametrized by a matrix $\mathbf{C}$  such that $c_{ij} = 1$ if the vertices $i$ and $j$ belong to the same community, and $c_{ij} = 0$ otherwise) and search for the optimal partition. The latter is the one that maximizes the modularity function which, for binary networks, is defined as
\begin{equation}
Q(\mathbf{C}) \equiv \frac{1}{k_{tot}}\sum_{i,j}\Big[a_{ij}-\frac{k_i k_j}{k_{tot}}\Big]c_{ij}
\label{eq_4}
\end{equation}
where eq.\eqref{eq_2} appears explicitly as a null model for $a_{ij}$. 
For weighted networks, a similar expression involving eq.\eqref{eq_3} applies.
Other important examples where eq.\eqref{eq_2} is used are the characterization of the connected components of
networks \cite{Chung et al 2002b}, the average distance among vertices \cite{Chung et al 2002a}, and more in general the theoretical study of percolation \cite{Newman 2010} (characterizing the system's
robustness under the failure of nodes and/or links) and other dynamical processes \cite{Barrat 2008} on networks. 

Due to the important role that these equations play in many applications, it is remarkable that the literature puts very little emphasis on the fact that eqs.\eqref{eq_2} and \eqref{eq_3} are valid only under strict conditions that, for most real networks, are strongly violated. 
It is evident that eq.\eqref{eq_2} represents a probability
only if the largest degree $k_{max}$ in the network does not exceed the so-called ``structural cut-off'' $k_c\equiv \sqrt{k_{tot}}$ \cite{Boguna et al 2004}, i.e. if
\begin{equation}
k_{max}<\sqrt{k_{tot}}
\label{eq_6}
\end{equation}
Obviously, the above condition sets an upper bound for the allowed heterogeneity of the degrees, since both $k_{max}$ and $k_{tot}$ are determined by the same degree distribution. Unfortunately, as we discuss below, it has been shown that $k_{max}$ strongly exceeds $k_{c}$ in most real-world networks, making eq.\eqref{eq_2} ill-defined. 

It should be noted that in principle the knowledge of $p_{ij}$ allows one to sample networks from the canonical ensemble very easily, by running over all pairs of nodes and connecting them with the appropriate probability.
However, the fact that $p_{ij}\gg 1$ when $k_{max}\gg k_{c}$ makes such probability useless for sampling purposes.
This is why, despite their conceptual simplicity, general algorithms to sample canonical ensembles of networks have not been implemented so far, and the emphasis has remained on microcanonical approaches. 

\subsection{The `strong heterogeneity regime' challenging most algorithms\label{sec:problems}}
Equations \eqref{eq_1}, \eqref{eq_0} and \eqref{eq_6}, along with our discussion above, show that most methods run into problems when the heterogeneity of the network is too pronounced: 
strongly heterogeneous networks elude most microcanonical  and canonical approaches proposed so far.
Unfortunately, networks in this extreme regime are known to be ubiquitous, and represent the rule rather than the exception. A simple way to prove this is by directly checking whether the largest degree exceeds the structural cut-off $k_c$. 
As Maslov et al. first noticed \cite{MS}, in real
networks $k_c$ is strongly and systematically exceeded: for instance, for the Internet $k_{max} = 1458$ and
$k_c\approx 159$, which means that the structural cut-off is exceeded ten-fold. 
Consequently, if eq. \eqref{eq_2} were applied
to the two vertices with largest degree, the resulting connection `probability' would be $p_{ij} = 43.5$, i.e. more
than 40 times larger than any reasonable estimate for a probability. 
We also note that, when inserted into eq.\eqref{eq_4}, this value of $p_{ij}$ would produce, in the
summation, a single term 40 times larger than any other `regular' (i.e. of order unity) term, thus significantly biasing the
community detection problem. To the best of our knowledge, a study of the entity of this bias has never been performed.

The Internet is not a special case, and similar results are found in the majority of real networks, making the problem entirely general. 
To see this, it is enough to exploit the fact that most real networks have a power-law degree distribution of the form $P(k)\sim k^{-\gamma}$ with exponent in the range $2 <\gamma < 3$. For these networks, the average degree $\overline{k}=k_{tot}/N$ is finite but the second moment $\overline{k^2}$ diverges.
Therefore the structural cut-off scales as $k_c\sim N^{1/2}$ \cite{Boguna et al 2004}, which means that eqs. \eqref{eq_0} and \eqref{eq_6} coincide. By contrast, Extreme
Value Theory shows that the largest degree scales as $k_{max}\sim N^{1/(\gamma -1)}$ \cite{Boguna et al 2004}. This implies that the ratio $k_{max}/k_c$ diverges for large networks, i.e. the largest degree is infinitely larger than the allowed cut-off value. 
Unfortunately, many results and approaches that have been obtained by assuming $k_{max}<k_c$ are
naively extended to real networks where, in most of the cases, $k_{max}\gg k_c$. 
Therefore, although this might appear as an exaggerated claim, most analyses of real-world networks (including community
detection) that have been carried out so far have relied on incorrect expressions, and have been systematically affected by an uncontrolled bias.

In theoretical and computational models of networks, the problem is normally circumvented by enforcing the
condition $k_{max}<k_c$ explicitly, e.g. by considering a truncated power-law distribution. This procedure is usually justified with the expectation that the inequality $k_{max}<k_c$ should hold for sparse networks where the average degree does not grow
with $N$, as in most real networks \cite{Newman et al 2001,Catanzaro et al 2005}. This interpretation of the role of sparsity is however misleading, since in real scale-free networks with $2 < \gamma < 3$ the average degree is finite irrespective of the presence of the cut-off. This makes those networks sparse even without assuming a truncation in the degree distribution. 
As a matter of fact, as clear from  the example above, real
networks systematically violate the cut-off value, and are therefore `strongly heterogeneous', even if sparse. 
By the way, the fact that a high density is not the origin of the breakdown of the available approaches should be clear by considering that dense but homogeneous networks (including the densest of all, i.e. the complete graph) are such that $k_{max}<k_c$ and are therefore correctly described by eq.\eqref{eq_2}, just like sparse
homogeneous networks. This confirms that the problem is in fact due to \emph{strong heterogeneity} and not to high density.

The above arguments can be extended to other ensembles of networks with different constraints. The general conclusion is that, since real-world networks are generally strongly heterogeneous, the available approaches either break down or become computationally demanding. Moreover, it is difficult to generalize the available knowledge to modified constraints and different types of graphs.

\section{The ``Max \& Sam'' method\label{sec:maxsam}}
In what follows, building on a series of recent results  characterizing several canonical ensembles of networks \cite{mymethod,mygrandcanonical,mywreciprocity,mybosefermi,myenhanced}, we introduce a unified approach to sample these ensembles in a fast, unbiased and efficient way.
In our approach, the functional form of the probability of each graph in the ensemble is derived by maximizing Shannon's entropy \cite{newman_expo} (thus ensuring that the sampling is unbiased), and the numerical coefficients of this probability are derived by maximizing the probability (i.e. the likelihood) itself \cite{mymethod}.
Since this double maximization is the core of our approach, we call our method the ``Maximize and Sample'' (``Max \& Sam'' for short) method.
We also provide a code implementing all our sampling algorithms (see Appendix).

We will consider canonical ensembles of binary graphs with given degree sequence (both undirected \cite{mymethod,myPRE1} and directed \cite{mymethod,myPRE1,mynull}), of weighted networks with given strength sequence (both undirected \cite{mymethod,myPRE2} and directed \cite{mymethod,myPRE2,mynull,mywreciprocity}), of directed networks with given reciprocity structure (both binary \cite{mygrandcanonical,mymotifs} and weighted \cite{mywreciprocity}), and of weighted networks with given combined strength sequence and degree sequence \cite{mybosefermi,myenhanced,myextensive}.
In all these cases, that have been treated only separately so far, we implement an explicit sampling protocol based on the exact result that the probability of the entire network always factorizes as a product of dyadic probabilities over pairs of nodes.
This ensures that the computational complexity of our sampling method is always $O(N^2)$ in all cases considered here, irrespective of the level of heterogeneity of the real-world network being randomized.
Therefore our method does not suffer from the limitations of the other methods discussed in sec. \ref{sec:previous}: it is efficient and unbiased even for strongly heterogeneous networks.

It should be noted that, while most microcanonical algorithms require as input the entire adjacency matrix of the observed graph (see sec. \ref{sec:micro}), our canonical approach requires only the empirical values of the constraints (e.g. the degree sequence).
At a theoretical level, this desirable property restores the expectation that such constraints should be the sufficient statistics of the problem. 
At a practical level, it enormously simplifies the data requirements of the sampling process.
For instance, if the sampling is needed in order to reconstruct an unknown network from partial node-specific information (e.g. to generate a collection of likely graphs consistent with an observed degree and/or strength sequence), then most microcanonical algorithms cannot be applied, while canonical ones can reconstruct the network to a high degree of accuracy \cite{myenhanced}.

\subsection{Binary undirected graphs with given degree sequence\label{sec:UBCM}}
Let us start by considering binary, undirected networks (BUNs). 
A generic BUN is uniquely specified by its binary adjacency matrix $\mathbf{A}$. 
The particular matrix corresponding to the observed graph that we want to randomize will be denoted by $\mathbf{A}^*$. 
As we mentioned, the simplest non-trivial constraint is the degree sequence, $\{k_i\}_{i=1}^N$ (where $k_i\equiv\sum_{j} a_{ij}$ is the degree of node $i$), defining the UBCM. 

In our approach, the canonical ensemble of BUNs is the set of networks with the same number of nodes, $N$, of the observed graph and a number of (undirected) links varying from zero to the maximum value $\frac{N(N-1)}{2}$. Appropriate probability distributions on this ensemble can be fully determined by maximizing, in sequence,  Shannon's entropy (under the chosen constraints) and the likelihood function, as already pointed out in \cite{mymethod}.
The result of the entropy maximization \cite{newman_expo,mymethod} is that the graph probability factorizes as 
\begin{equation}
P(\mathbf{A}|\mathbf{x})=\prod_{i}\prod_{j<i}p_{ij}^{a_{ij}}(1-p_{ij})^{1-a_{ij}}\label{cm}
\end{equation}
where $p_{ij}\equiv\frac{x_ix_j}{1+x_ix_j}$.
The vector $\mathbf{x}$ of $N$ unknown parameters (or `hidden variables') is to be determined either by maximizing the log-likelihood function 
\begin{eqnarray}
\lambda(\mathbf{x})&\equiv&\ln P(\mathbf{A}^*|\mathbf{x})=\\
&=&\sum_i k_i(\mathbf{A}^*) \ln x_i - \sum_i\sum_{j<i} \ln (1+x_ix_j)\nonumber
\label{CM}
\end{eqnarray} 
or, equivalently, by solving the following system of $N$ equations (corresponding to the requirement that the gradient of the log-likelihood vanishes) \cite{mymethod}:
\begin{equation}
\langle k_i \rangle=\sum_{j \neq i} \frac{x_ix_j}{1+x_ix_j}=k_i(\mathbf{A}^*) \quad\forall i\label{CMeq}
\end{equation}
where $k_i(\mathbf{A}^*)$ is the observed degree of vertex $i$ and $\langle k_i\rangle$ indicates its ensemble average.
In both cases, the parameters $\mathbf{x}$ vary in the region defined by $ x_i\ge 0$ for all $i$  \cite{mymethod}.

From eq.\eqref{CM} it is evident that only the observed values of the chosen constraints (the \emph{sufficient statistics} of the problem) are needed in order to obtain the numerical values of the unknowns (the empirical degree sequence fixes the value of $\mathbf{x}$, which in turn fix the value of all the probabilities $\{ p_{ij}\}$).
In any case, for the sake of clarity, in the code we allow the user to choose the preferred input-form (a matrix, a list of edges, a vector of constraints). This applies to all the models described in this paper and implemented in the code.
\begin{figure*}[t!]
\includegraphics[width=0.8\textwidth]{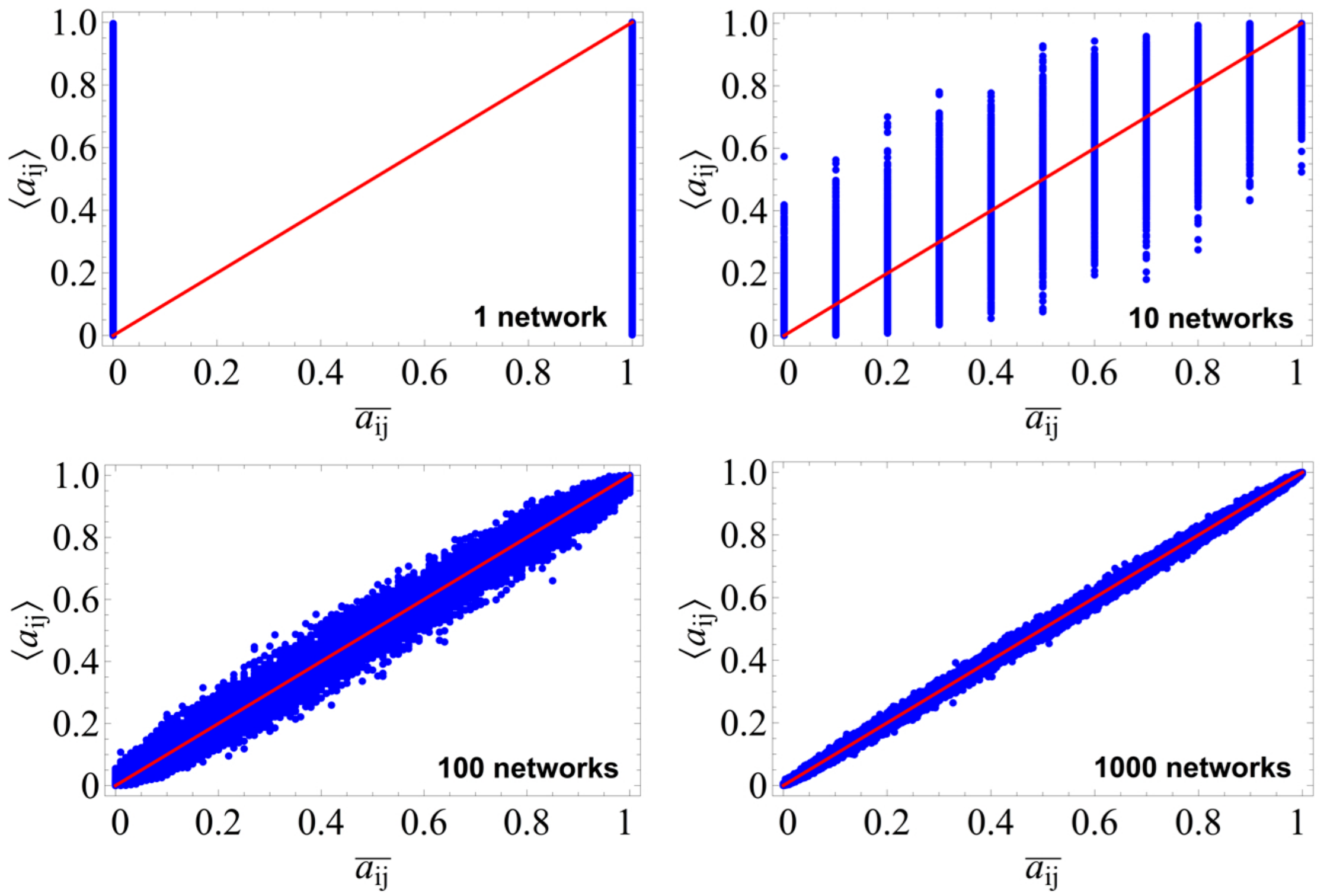}
\caption{Sampling binary undirected networks with given degree sequence (Undirected Binary Configuration Model). The example shown is the binary network of liquidity reserves exchanges between Italian banks in 1999 \cite{Bankdata} (N = 215).  
The four panels show the convergence of the sample average $\overline{a_{ij}}$ of each entry of the adjacency matrix to its exact canonical expectation $\langle a_{ij}\rangle$, for 1 (top left), 10 (top right), 100 (bottom left) and 1000 (bottom right) sampled networks. The identity line is shown in red.}
\label{bun_a2}
\end{figure*}

Note that the above form of $p_{ij}$ represents the exact expression that should be used in place of eq.\eqref{eq_2}.
This reveals the highly non-linear and non-local character of the interdependencies among vertices in the UBCM: in random networks with given degree sequence, the correct connection probability $p_{ij}$ is a function of the degrees of \emph{all vertices} of the network, and not just of the end-point degrees as in eq.\eqref{eq_2}. 
Only when the degrees are `weakly heterogeneous' (mathematically, this happens when $x_i x_j\ll 1$ for all pairs of vertices, which implies $p_{ij}\approx x_i x_j$), these structural interdependencies become
approximately local. 
Note that, in the literature, this is improperly called the ``sparse graph'' limit \cite{newman_expo}, while, as we discussed in sec.\ref{sec:problems}, what defines this limit is a low level of heterogeneity, and not sparsity.

\begin{figure*}[t!]
\includegraphics[width=0.8\textwidth]{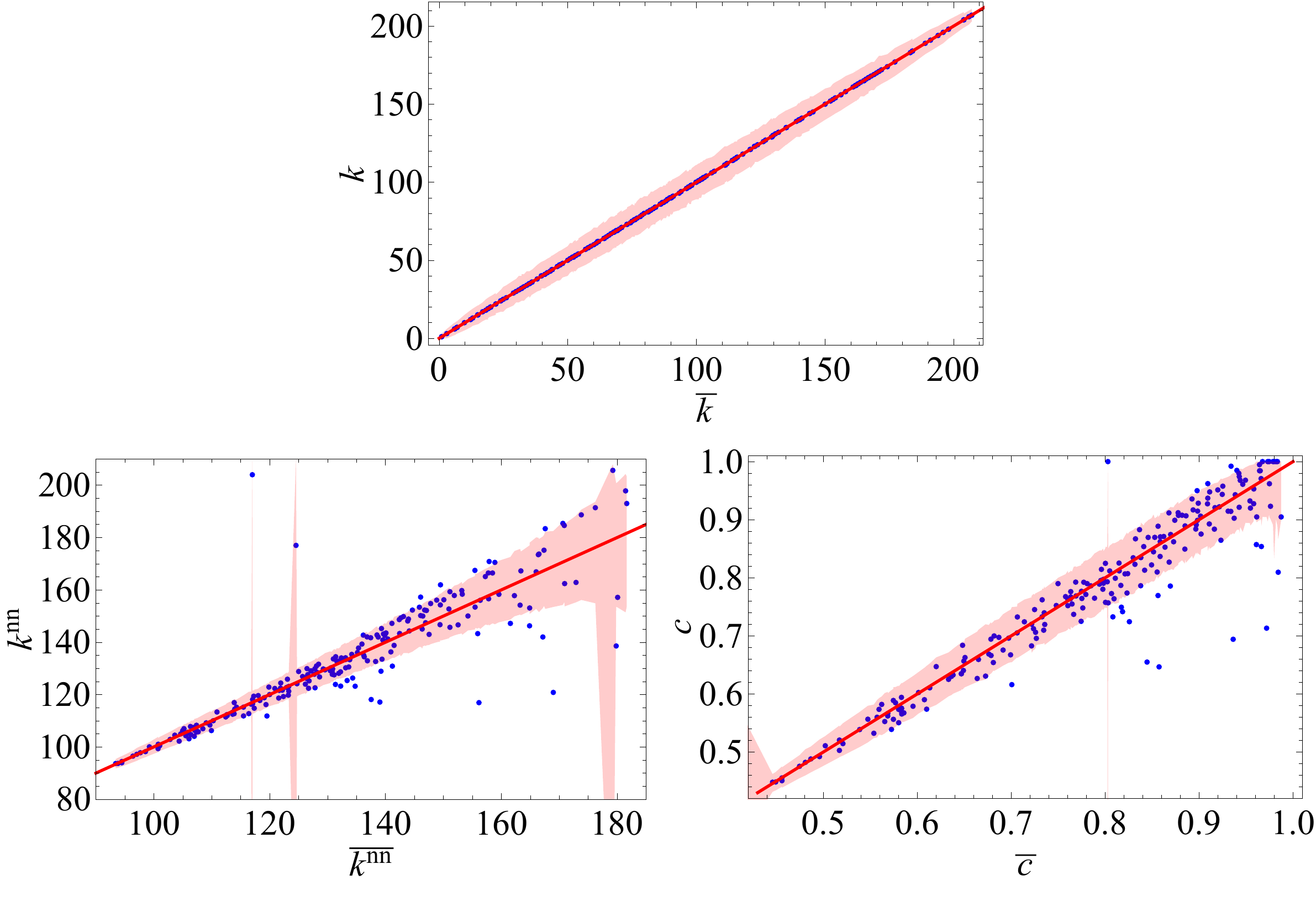}
\caption{Sampling binary undirected networks with given degree sequence (Undirected Binary Configuration Model). The example shown is the binary network of liquidity reserves exchanges between Italian banks in 1999 \cite{Bankdata} (N = 215). 
The three panels show, for each node in the network, the comparison between the observed value and the sample average of the (constrained) degree (top), the (unconstrained) ANND (bottom left) and the (unconstrained) clustering coefficient (bottom right), for 1000 sampled networks. The 95\% confidence intervals of the distribution of the sampled quantities is shown in pink for each node.}
\label{bun_top2}
\end{figure*}

Unlike eq.\eqref{eq_2}, the $p_{ij}$ considered here always represents a proper probability ranging between 0 and 1, irrespective of the heterogeneity of the network.
This implies that eq.\eqref{cm} provides us with a recipe to sample the canonical ensemble of BUNs under the UBCM.
After the unknown parameters have been found, they can be put back into eq.\eqref{cm} to obtain the probability to correctly sample any graph $\mathbf{A}$ from the ensemble.
The key simplification allowing this in practice is the fact that the graph probability is factorized, so that a single graph can be sampled stochastically by sequentially running over each pair of nodes $i,j$ and implementing a Bernoulli trial (whose elementary events are $a_{ij}=0$, with probability $1-p_{ij}$, and $a_{ij}=1$, with probability $p_{ij}$). This process can be repeated to generate as many configurations as desired. 
Note that sampling each network has complexity $O(N^2)$, and that the time required to preliminarily solve the system of coupled equations to find the unknown parameters $\mathbf{x}$ is independent on how many random networks are sampled and on the heterogeneity of the network.
Thus this algorithm is always more efficient than the corresponding microcanonical ones described in sec.\ref{sec:micro}.

In fig. \ref{bun_a2} we show an application of this procedure to the network of liquidity reserves exchanges between Italian banks in 1999 \cite{Bankdata}. 
For an increasing number of sampled graphs, we show the convergence of the sample average $\overline{a_{ij}}$ of each entry of the adjacency matrix to its exact canonical expectation $\langle a_{ij}\rangle$, analytically determined after solving the likelihood equations.
This preliminary check is useful to establish that, in this case, generating 1000 networks (bottom right) is enough to reach a high level of accuracy. 
If needed, the accuracy can be quantified rigorously (e.g. in terms of the maximum width around the identity line) and arbitrarily improved by increasing the number of sampled matrices.
Note that this important check is impossible in microcanonical approaches, where the exact value of the target probability is unknown.

We then select the sample of 1000 networks and confirm (see the top panel of fig. \ref{bun_top2}) that the imposed constraints (the observed degrees of all nodes) are very well reproduced by the sample average, and that the confidence intervals are narrowly spread around the identity line. 
This is an important test of the accuracy of our sampling procedure. Again, the accuracy can be improved by increasing the number of sampled matrices if needed.

After this preliminary check, the sample can be used to compare the expected and observed values of higher-order properties of the network. Note that in this case we do not require (or expect) that these (unconstrained) higher-order properties are correctly reproduced by the null model. 
The entity of the deviations of the real network from the null model depends on the particular example considered, and the characterization of these deviations is precisely the reason why a method to sample random networks from the appropriate ensemble is needed in the first place.
In the bottom panels of fig. \ref{bun_top2} we compare the observed value of two quantities of interest with their arithmetic mean over the sample. 
The two quantities are the average nearest neighbors degree (ANND), $k_i^{nn}=\frac{\sum_ja_{ij}k_j}{k_i}$, and the clustering coefficient, $c_i=\frac{\sum_{j,\:k}a_{ij}a_{jk}a_{ki}}{k_i(k_i-1)}$ of each vertex. 

Note that, since our sampling method is unbiased, the arithmetic mean over the sample automatically weighs the configurations according to their correct probability. 
In this particular case, we find that the null model reproduces the observed network very well, which means that the degree sequence effectively explains (or rather generates) the two empirical higher-order patterns that we have considered. 
This is consistent with other studies \cite{mymethod,myPRE1,myPRE2}, but not true in general for other networks or other constraints, as we show later on. 
From the bottom panels of fig. \ref{bun_top2} we also note that the confidence intervals highlight a non-obvious feature: the fact that the few points further away from the identity line turn out to be actually within (or at the border of) the chosen confidence intervals, while several points closer to the identity are instead found to be much more distant from the confidence intervals, and thus in an unexpectedly stronger disagreement with the null model.
These counter-intuitive insights cannot be derived from the analysis of the expected values alone, e.g. using expressions like eq.\eqref{eq_2} or similar.

\begin{figure}[t]
\includegraphics[width=0.48\textwidth]{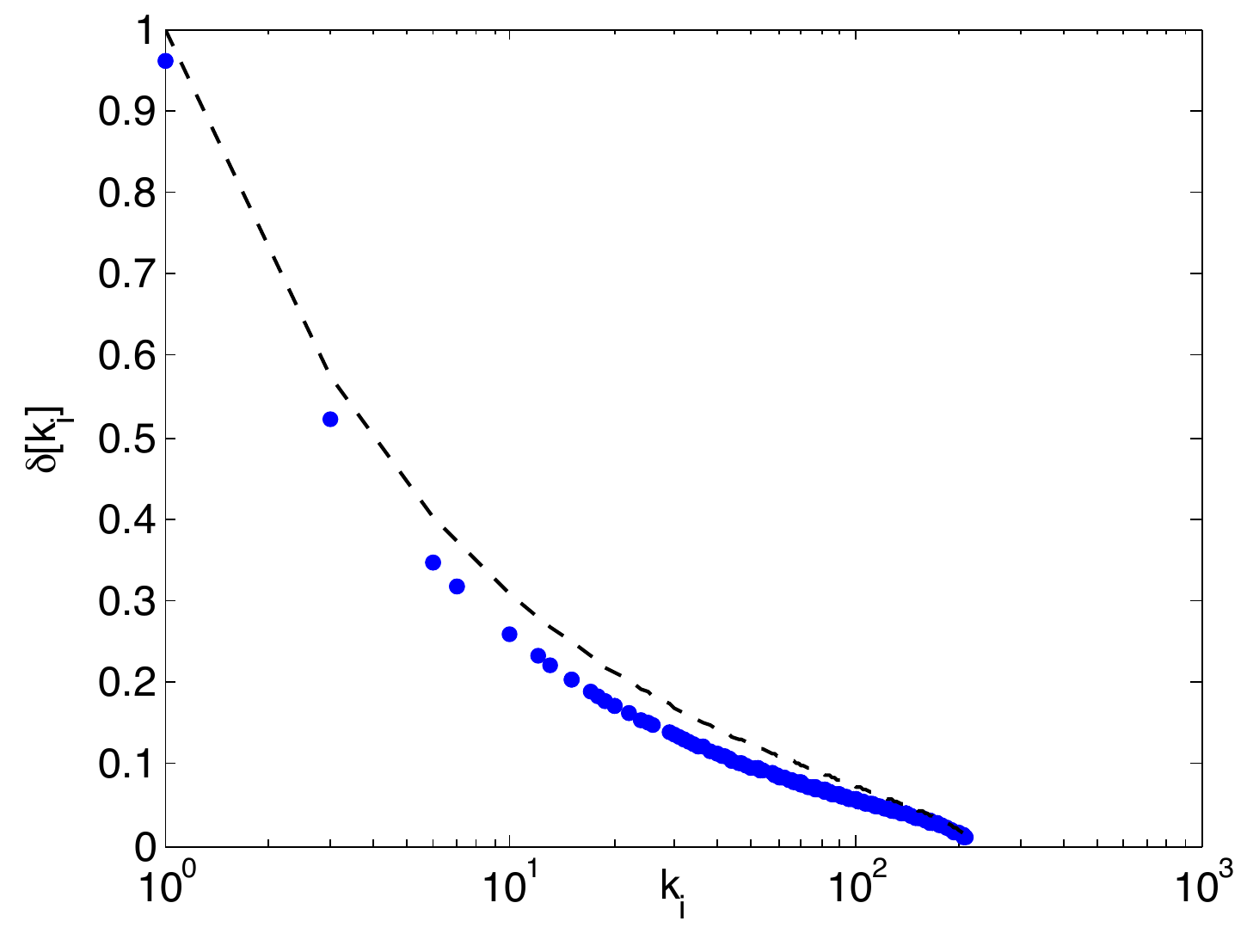}
\caption{Coefficient of variation $\delta[k_i]$ as a function of the degree $k_i$ for each node of the binary network of liquidity reserves exchanges between Italian banks in 1999 \cite{Bankdata} ($N$ = 215). 
The blue points are the exact values in eq.(\ref{eq:fluUBCM}), while the dashed curve is the upper bound in eq.(\ref{eq:boundUBCM}). The lower bound is the abscissa $\delta[k_i]=0$.}
\label{fig:fluUBCM}
\end{figure}

We now calculate the fluctuations of the constraints explicitly.
We start by calculating the ensemble variance of each degree $k_i$, defined as $\sigma^2[k_i]\equiv\langle k_i^2\rangle-\langle k_i\rangle^2$. In the microcanonical ensemble, one obviously has $\sigma^2[k_i]=0$.
In the canonical ensemble, the independence of pairs of nodes implies that the variance of the sum $\sum_{j\ne i}a_{ij}$ coincides with the sum of the variances of its terms, i.e.
\begin{eqnarray}
\sigma^2[k_i]&=&\sum_{j\ne i}\sigma^2[a_{ij}]=\sum_{j\ne i}(\langle a^2_{ij}\rangle-\langle a_{ij}\rangle^2)\nonumber\\
&=&\sum_{j\ne i}p_{ij}(1-p_{ij})=k_i-\sum_{j\ne i}p_{ij}^2.\label{eq:varUBCM}
\end{eqnarray}
Then, the canonical relative fluctuations can be measured in terms of the so-called \emph{coefficient of variation}, which we conveniently express in the form
\begin{equation}
\delta[k_i]\equiv\frac{\sigma[k_i]}{k_i}=\sqrt{\frac{1}{k_i}-\frac{\sum_{j\ne i}p_{ij}^2}{(\sum_{j\ne i}p_{ij})^2}},
\label{eq:fluUBCM}
\end{equation}
where we have restricted ourselves to the case $k_i>0$\footnote{The case $k_i=0$ also implies $\sigma[k_i]=0$ and leads to an indeterminate form for $\delta[k_i]$. However this case is uninteresting since each isolated node $i$ remains isolated across the entire ensemble ($p_{ij}=0$ $\forall j$) and can be safely removed without loss of generality.}.
A plot of $\delta[k_i]$ as a function of $k_i$ for the interbank network considered above is shown in fig. \ref{fig:fluUBCM}.
We find that the relative fluctuations vanish for vertices with large degree, while they are very large for vertices with moderate and small degree. In particular, $\delta[k_i]\approx 1$ when $k_i=1$.

In general, we note that the term ${\sum_{j\ne i}p_{ij}^2}/{(\sum_{j\ne i}p_{ij})^2}$ in eq.(\ref{eq:fluUBCM}) is a \emph{participation ratio} \footnote{Strictly speaking, it is the inverse of a so-called \emph{inverse participation ratio}, but we avoid the use of `inverse' twice.}, measuring the inverse of the effective number of equally important terms in the sum ${\sum_{j\ne i}p_{ij}}$: in particular, it equals $1$ if and only if there is only one nonzero term (complete concentration), while it equals $(N-1)^{-1}$ if and only if there are $N-1$ identical terms (complete homogeneity), i.e. $p_{ij}=k_i/(N-1)$ for all $j\ne i$. 
Since these are the two extreme bounds for a participation ratio, and since in the case of complete concentration we also have $k_i=1$, we conclude that the bounds for $\delta[k_i]$ are
\begin{equation}
0\le\delta[k_i]\le\sqrt{\frac{1}{k_i}-\frac{1}{N-1}}.
\label{eq:boundUBCM}
\end{equation}
The resulting allowed region for $\delta[k_i]$ is the one comprised between the abscissa and the dashed line in  fig.\ref{fig:fluUBCM}. We find that the realized trend is  close to the upper bound. This suggests that the maximum-entropy nature of our algorithm produces almost maximally homogeneous terms in the sum ${\sum_{j\ne i}p_{ij}}$, i.e. no particular subset of vertices is preferred as canditate partners for $i$, the only preference being obviously given (as a consequence of the explicit form of $p_{ij}$ in terms of $x_i$ and $x_j$) to vertices with larger degree. 

Since the degree distribution of most real-world networks is such that the average degree remains finite even when the size of the network becomes very large, the above results suggest that, unlike most physical systems, the microcanonical and canonical ensembles defined by the UBCM are \emph{not} equivalent in the `thermodynamic' limit $N\to\infty$.
While eq.(\ref{eq:boundUBCM}) shows that values closer to the lower bound $\delta[k_i]=0$ can be in principle achieved, the maximization of the entropy appears to push the ensemble towards the opposite upper bound where the equivalence of the microcanonical and canonical ensembles is maximally violated. 
On the other hand, one might in principle construct synthetic networks with sufficiently large degrees, such that the canonical fluctuations are arbitrarily small and the two ensembles arbitrarily close.

\subsection{Binary directed graphs with given in-degree and out-degree sequences\label{sec:DBCM}}
For binary directed networks (BDNs), the adjacency matrix $\mathbf{A}$ is (in general) not symmetric, and each node $i$ is characterized by two degrees: the out-degree $k_i^{out}\equiv\sum_j a_{ij}$ and the in-degree $k_i^{in}\equiv\sum_j a_{ji}$.
The \emph{Directed Binary Configuration Model} (DBCM), which is the directed version of the UBCM, is defined as the ensemble of BDNs with given out-degree sequence $\{k_i^{out}\}_{i=1}^N$ and in-degree sequence $\{k_i^{in}\}_{i=1}^N$.

At a canonical level, the DBCM is defined on the ensemble of all BDNs with $N$ vertices and a number of links ranging from 0 to $N(N-1)$. 
Equation \eqref{cm} still applies, but now with ``$j<i$'' replaced by ``$j\ne i$'' and $p_{ij}=\frac{x_iy_j}{1+x_iy_j}$, where the $2N$ parameters $\mathbf{x}$ and $\mathbf{y}$ are determined by either maximizing the log-likelihood function \cite{mymethod} 
\begin{eqnarray}
\lambda(\mathbf{x},\mathbf{y})&\equiv&\ln P(\mathbf{A}^*|\mathbf{x},\mathbf{y})=\\
&=&\sum_i \big[k_i^{out}  (\mathbf{A}^*) \ln x_i + k_i^{in}(\mathbf{A}^*)\ln y_i\big]\nonumber\\
&-&\sum_i\sum_{j \neq i} \ln (1+x_iy_j)\nonumber
\end{eqnarray}
(where $\mathbf{A}^*$ is the real network) or, equivalently, by solving the system of $2N$ equations \cite{mymethod}
\begin{eqnarray}  
\langle k_i^{out} \rangle&=&\sum_{j \neq i} \dfrac{x_iy_j}{1+x_iy_j}=k_i^{out}(\mathbf{A}^*)\quad \forall i \\
\langle k_i^{in} \rangle&=&\sum_{j \neq i} \dfrac{x_jy_i}{1+x_jy_i}=k_i^{in}(\mathbf{A}^*)\quad \forall i.
\end{eqnarray}
The parameters $\mathbf{x}$ and $\mathbf{y}$ vary in the region defined by $ x_i\ge 0$ and $y_i\ge 0$  for all $i$ respectively \cite{mymethod}.

The ensemble can be efficiently sampled by considering each pair of vertices \emph{twice}, and using (say) $p_{ij}$ and $p_{ji}$ to draw directed links in the two directions (these two events being statistically independent). Since this is a straightforward extension of the UBCM, we do not consider any specific example to illustrate the DBCM.
However, the related algorithm has been implemented in the code (see Appendix).

We conclude the discussion of this ensemble with the calculation of the canonical fluctuations. In analogy with eqs.(\ref{eq:varUBCM}) and (\ref{eq:fluUBCM}), the variances of $k^{out}_i$ and $k^{in}_i$ are given by 
\begin{eqnarray}
\sigma^2[k^{out}_{i}]&=& \sum_{j \ne i} p_{ij}(1-p_{ij})=k^{out}_i-\sum_{j \ne i} p^2_{ij},\label{eq:varDBCM1}\\
\sigma^2[k^{in}_{i}]&=& \sum_{j \ne i} p_{ji}(1-p_{ji})=k^{in}_i-\sum_{j \ne i} p^2_{ji}.\label{eq:varDBCM2}
\end{eqnarray}
For $k^{out}_i>0$ and $k^{in}_i>0$, the relative fluctuations are
\begin{eqnarray}
\delta[k^{out}_i]&\equiv&\frac{\sigma[k^{out}_i]}{k^{out}_i}=\sqrt{\frac{1}{k^{out}_i}-\frac{\sum_{j\ne i}p_{ij}^2}{(\sum_{j\ne i}p_{ij})^2}},\label{eq:fluDBCM1}\\
\delta[k^{in}_i]&\equiv&\frac{\sigma[k^{in}_i]}{k^{in}_i}=\sqrt{\frac{1}{k^{in}_i}-\frac{\sum_{j\ne i}p_{ji}^2}{(\sum_{j\ne i}p_{ji})^2}}.\label{eq:fluDBCM2}
\end{eqnarray}
The above quantities still involve participation ratios defined by the connection probabilities.
For the bounds of $\delta[k^{out}_i]$ and $\delta[k^{in}_i]$, considerations similar to those leading to eq.(\ref{eq:boundUBCM}) apply here.

\subsection{Binary directed graphs with given degree sequences and reciprocity structure\label{sec:RBCM}}
A more constrained null model, the \emph{Reciprocal Binary Configuration Model} (RBCM), can be defined for BDNs
by enforcing, in addition to the two directed degree sequences considered above, the whole local reciprocity structure of the network \cite{mymethod,mygrandcanonical,mymotifs}.
This is equivalence to the specification of the three  degree sequences defined as the vector of the numbers of non-reciprocated outgoing links, $\{k_i^{\rightarrow}\}_{i=1}^N$, the vector of the numbers of non-reciprocated incoming links, $\{k_i^{\leftarrow}\}_{i=1}^N$, and the vector of the numbers of reciprocated links, $\{k_i^{\leftrightarrow}\}_{i=1}^N$ \cite{mymethod,mygrandcanonical,mymotifs}.
These numbers are defined as $k_i^{\rightarrow}\equiv\sum_j a_{ij}(1-a_{ji})$, $k_i^{\leftarrow}\equiv\sum_j a_{ji}(1-a_{ij})$, and $k_i^{\leftrightarrow}\equiv\sum_j a_{ij}a_{ji}$ respectively \cite{mygrandcanonical,mymotifs}.

The RBCM is of crucial importance when analysing higher-order patterns that exist beyond the dyadic level in directed networks.
The most important example is that of \emph{triadic motifs} \cite{motifs,mymotifs,Squartini et al 2013}, i.e. patterns of connectivity (involving triples of nodes) that are statistically over- or under-represented with respect to a null model where the observed degree sequences and reciprocity structure are preserved (i.e. the RBCM).
Note that in this case no approximate canonical expression similar to eq.\eqref{eq_2} exists, therefore the null model is usually implemented microcanonically using a generalization of the LRA that we have discussed in sec. \ref{sec:micro}.
Conceptually, this procedure suffers from the same problem of bias as the simpler procedures used to implement the UBCM and the DBCM through the LRA \cite{Artzy-Randrup
et al 2005,coolen2,coolen}. To our knowledge, in this case no correction analogous to that proposed in ref. \cite{coolen} has been developed in order to restore uniformity.

In our ``Max \& Sam'' approach, we exploit known analytical results \cite{mymethod,mygrandcanonical,mymotifs} showing that the probability of each graph $\mathbf{A}$ in the RBCM is
\begin{equation}
P(\mathbf{A}|\mathbf{x},\mathbf{y},\mathbf{z})=\prod_i\prod_{j<i}(p_{ij}^{\rightarrow})^{a_{ij}^{\rightarrow}}(p_{ij}^{\leftarrow})^{a_{ij}^{\leftarrow}}(p_{ij}^{\leftrightarrow})^{a_{ij}^{\leftrightarrow}}(p_{ij}^{\nleftrightarrow})^{a_{ij}^{\nleftrightarrow}},
\end{equation}
where $p_{ij}^{\rightarrow}\equiv\frac{x_iy_j}{1+x_iy_j+x_jy_i+z_iz_j}$, $p_{ij}^{\leftarrow}\equiv\frac{x_jy_i}{1+x_iy_j+x_jy_i+z_iz_j}$, $p_{ij}^{\leftrightarrow}\equiv\frac{z_iz_j}{1+x_iy_j+x_jy_i+z_iz_j}$ and $p_{ij}^{\nleftrightarrow}\equiv\frac{1}{1+x_iy_j+x_jy_i+z_iz_j}$ denote the probabilities of a single (non-reciprocated) link from $i$ to $j$, a single (non-reciprocated) link from $j$ to $i$, a double (reciprocated) link between $i$ and $j$, and no link at all respectively.
The above four possible events are mutually exclusive. The greatest difference with respect to the DBCM lies in the fact that the two links that can be drawn between the same two nodes are no longer independent.

The $3N$ unknown parameters, $\mathbf{x}$, $\mathbf{y}$ and $\mathbf{z}$, must be determined by either maximizing the log-likelihood \cite{mymethod}
\begin{eqnarray}  
\lambda(\mathbf{x},\mathbf{y},\mathbf{z})&\equiv&
\ln P(\mathbf{A}^*|\mathbf{x},\mathbf{y},\mathbf{z})=\\
&=&\sum_i \big[k_i^{\rightarrow}(\mathbf{A}^*) \ln x_i + k_i^{\leftarrow}(\mathbf{A}^*)\ln y_i\nonumber\\
&+& k_i^{\leftrightarrow}(\mathbf{A}^*) \ln z_i\big] 
-\sum_i\sum_{j<i} \ln (1\!+\!x_iy_j\!+\!x_jy_i\!+\!z_iz_j)\nonumber
\end{eqnarray}
or, equivalently, solving the $3N$ coupled equations  \cite{mymethod,mygrandcanonical,mymotifs}:
\begin{eqnarray}
\langle k_i^{\rightarrow} \rangle&=&\sum_{j \neq i} \dfrac{x_iy_j}{1+x_iy_j+x_jy_i+z_iz_j}=k_i^{\rightarrow}(\mathbf{A}^*)\quad\forall i \qquad\\
\langle k_i^{\leftarrow} \rangle&=&\sum_{j \neq i} \dfrac{x_jy_i}{1+x_iy_j+x_jy_i+z_iz_j}=k_i^{\leftarrow}(\mathbf{A}^*)\quad\forall i \\
\langle k_i^{\leftrightarrow} \rangle&=&\sum_{j \neq i} \dfrac{z_iz_j}{1+x_iy_j+x_jy_i+z_iz_j}=k_i^{\leftrightarrow}(\mathbf{A}^*)\quad\forall i.
\end{eqnarray}
The parameters $\mathbf{x}$, $\mathbf{y}$ and $\mathbf{z}$ vary in the region defined by $ x_i\ge 0$, $y_i\ge 0$ and $z_i\ge 0$ for all $i$ respectively \cite{mymethod}.

After the unknown parameters have been found, the four probabilities allow us to sample the ensemble correctly and very easily. In particular, we can consider each pair of vertices $i,j$ \emph{only once} and either draw a single link directed from $i$ to $j$ with probability $p_{ij}^{\rightarrow}$, draw a single link directed from $j$ to $i$ with probability $p_{ij}^{\leftarrow}$, draw two mutual links with probability $p_{ij}^{\leftrightarrow}$, or draw no link at all with probability $p_{ij}^{\nleftrightarrow}$. 
Note that, despite the increased number of constraints, the computational complexity is still $O(N^2)$.
As for the DBCM, we do not show a specific illustration of the RBCM, but the procedure described above has been fully coded in order to sample the relevant ensemble in a fast and  unbiased way (see Appendix).

Coming to the canonical fluctuations, in this ensemble eqs.(\ref{eq:varDBCM1}) and (\ref{eq:varDBCM2}) generalize to
\begin{eqnarray}
\sigma^2[k^{\to}_{i}]&=& \sum_{j \neq i} p^{\to}_{ij}(1-p^{\to}_{ij})=k^{\to}_i-\sum_{j \neq i} (p^{\to}_{ij})^2,\label{eq:varRBCM1}\\
\sigma^2[k^{\gets}_{i}]&=& \sum_{j \neq i} p^{\gets}_{ij}(1-p^{\gets}_{ij})=k^{\gets}_i-\sum_{j \neq i}(p^{\gets}_{ij})^2.\label{eq:varRBCM2}\\
\sigma^2[k^{\leftrightarrow}_{i}]&=& \sum_{j \neq i} p^{\leftrightarrow}_{ij}(1-p^{\leftrightarrow}_{ij})=k^{\leftrightarrow}_i-\sum_{j \neq i} (p^{\leftrightarrow}_{ij})^2.\label{eq:varRBCM3}
\end{eqnarray}
For $k^{\to}_i>0$, $k^{\gets}_i>0$ and $k^{\leftrightarrow}_i>0$, the relative fluctuations are
\begin{eqnarray}
\delta[k^{\to}_i]&\equiv&\frac{\sigma[k^{\to}_i]}{k^{\to}_i}=\sqrt{\frac{1}{k^{\to}_i}-\frac{\sum_{j\ne i}(p_{ij}^\to)^2}{(\sum_{j\ne i}p_{ij}^\to)^2}},\label{eq:fluRBCM1}\\
\delta[k^{\gets}_i]&\equiv&\frac{\sigma[k^{\gets}_i]}{k^{\gets}_i}=\sqrt{\frac{1}{k^{\gets}_i}-\frac{\sum_{j\ne i}(p_{ji}^\gets)^2}{(\sum_{j\ne i}p^\gets_{ji})^2}},\label{eq:fluRBCM2}\\
\delta[k^{\leftrightarrow}_i]&\equiv&\frac{\sigma[k^{\leftrightarrow}_i]}{k^{\leftrightarrow}_i}=\sqrt{\frac{1}{k^{\leftrightarrow}_i}-\frac{\sum_{j\ne i}(p_{ji}^\leftrightarrow)^2}{(\sum_{j\ne i}p^\leftrightarrow_{ji})^2}}.\label{eq:fluRBCM3}
\end{eqnarray}

Thus, in all the ensembles considered so far (which are defined in terms of \emph{purely binary} constraints), the squared relative fluctuation of each constraint always takes the form of the inverse of the value of the constraint itself, \emph{minus} the participation ratio of the corresponding probabilities.

\subsection{Weighted undirected networks with given strength sequence\label{sec:UWCM}}
Let us now consider weighted undirected networks (WUNs). Differently from the binary case, link weights can now range from zero to infinity by (without loss of generality) integer steps. The number of configurations in the canonical ensemble is therefore infinite. Still, enforcing node-specific constraints implies that a proper probability measure can be defined over the ensemble, such that the average value of any network property of interest is finite \cite{mymethod}.
A single graph in the ensemble is now specified by its (symmetric) weight matrix $\mathbf{W}$, where the entry $w_{ij}$ represents the integer weight of the link connecting nodes $i$ and $j$ ($w_{ij}=0$ means that no link is there).
We denote the particular real-world weighted network as $\mathbf{W}^*$.
Each vertex is characterized by its \emph{strength} $s_i=\sum_j w_{ij}$ representing the weighted analogue of the degree.

The weighted, undirected counterpart of the UBCM is the \emph{Undirected Weighted Configuration Model} (UWCM). The constraint defining it is the observed strength sequence, $\{s_i\}_{i=1}^N$.
Like its binary analogue, the UBCM is widely used in order to detect communities and other higher-order patterns in undirected weighted networks. 
However, most approaches \cite{Newman 2010} incorrectly assume that this model is characterized by eq. \eqref{eq_3}, which is instead only a highly simplified expression \cite{mymethod}.

In the canonical ensemble, the probability of each weighted network $\mathbf{W}$ is \cite{mymethod}
\begin{equation}
P(\mathbf{W}|\mathbf{x})=\prod_i\prod_{j<i}p_{ij}^{w_{ij}}(1-p_{ij})
\label{eq:Pw}
\end{equation}
where now $p_{ij}\equiv x_ix_j$, showing that the weights are drawn from geometric distributions \cite{mywrg}. 
As usual, the numerical values of the unknown parameters $\mathbf{x}$ are found by either maximizing the log-likelihood function 
\begin{eqnarray}
\lambda(\mathbf{x})&\equiv&
\ln P(\mathbf{W}^*|\mathbf{x})=\\
&=&\sum_i s_i(\mathbf{W}^*) \ln x_i +\sum_i\sum_{j<i} \ln (1-x_ix_j)\nonumber
\label{WCM}
\end{eqnarray}
or solving the system of $N$ equations:
\begin{eqnarray}
\langle s_i \rangle&=&\sum_{j \neq i} \dfrac{x_ix_j}{1-x_ix_j}=s_i(\mathbf{W}^*) \quad \forall i
\end{eqnarray}
In both approaches, now the parameters $\mathbf{x}$ vary in the region defined by the constraint $0\le x_i x_j<1$ for all $i,j$ \cite{mymethod}.

In this model, after finding the unknown parameters we can sample the canonical ensemble by drawing, for each pair of vertices $i$ and $j$, a link of weight $w$ with geometrically distributed probability $p_{ij}^{w}(1-p_{ij})$. Note that this correctly includes the case $w_{ij}=0$, occurring with probability $1-p_{ij}$, corresponding to the absence of a link.
Alternatively, using a procedure similar to that discussed in \cite{mywrg}, one can start with the disconnected vertices $i$ and $j$, draw a first link (of unit weight) with Bernoulli-distributed probability $p_{ij}$, and (only if this event is successful) place a second unit of weight on the same link, again with probability $p_{ij}$, and so on until a failure is first encountered. 
In this way, only repetitions of elementary Bernoulli trials are involved, a feature that can sometimes be convenient for coding purposes (e.g. if only uniformly random number generators need to be used).
After all pairs of vertices have been considered and a single weighted network has been sampled, the process can be repeated until the desired number of networks is sampled.

\begin{figure*}[t!]
\includegraphics[width=0.8\textwidth]{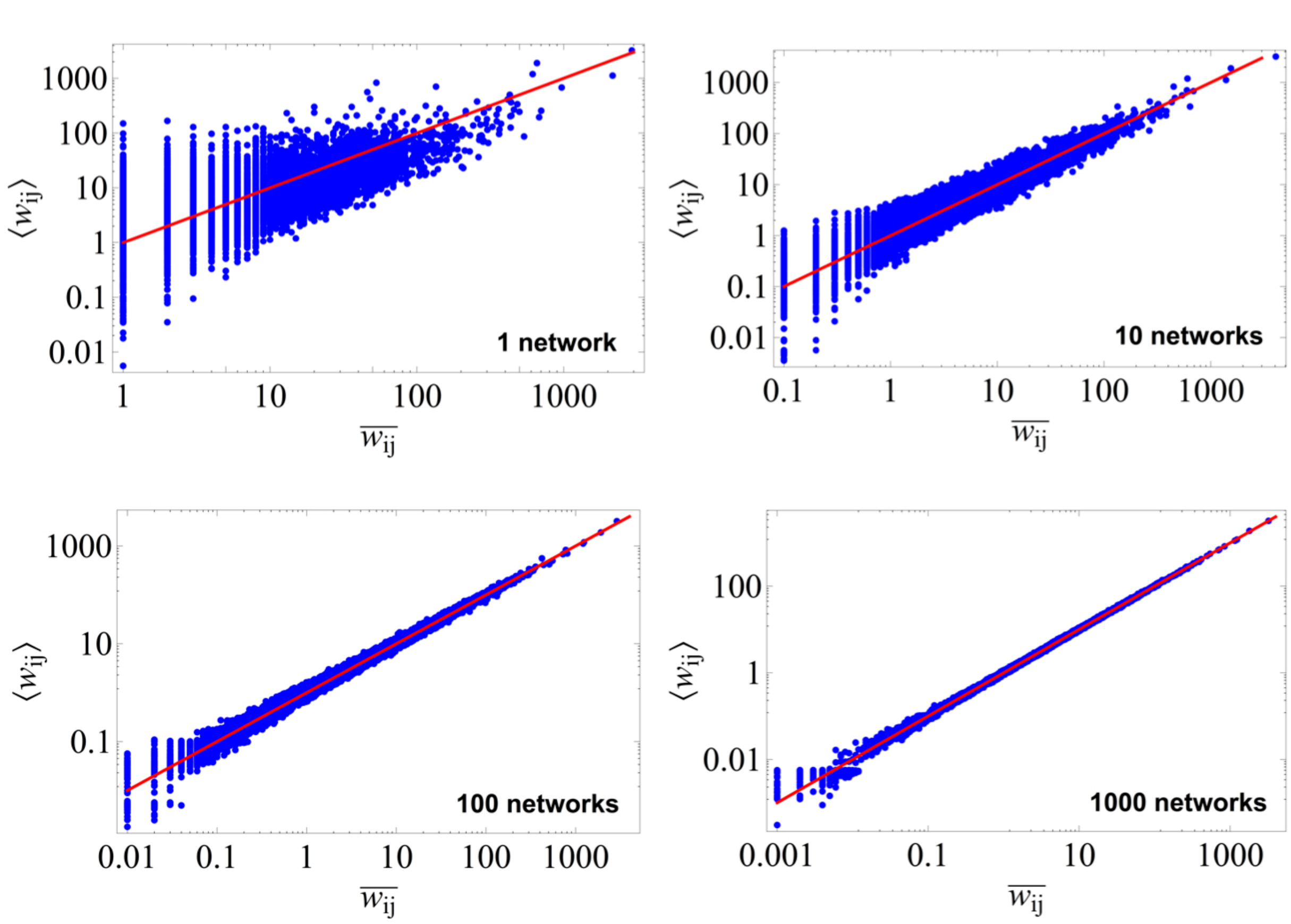}
\caption{Sampling weighted undirected networks with given strength sequence (Undirected Weighted Configuration Model). The example shown is the weighted network of liquidity reserves exchanges between Italian banks in 1999 \cite{Bankdata} (N = 215). The four panels show the convergence of the sample average $\overline{w_{ij}}$ of each entry of the weight matrix to its exact canonical expectation $\langle w_{ij}\rangle$, for 1 (top left), 10 (top right), 100 (bottom left) and 1000 (bottom right) sampled networks. The identity line is shown in red.}
\label{wun_w2}
\end{figure*}
\begin{figure*}[t!]
\includegraphics[width=0.8\textwidth]{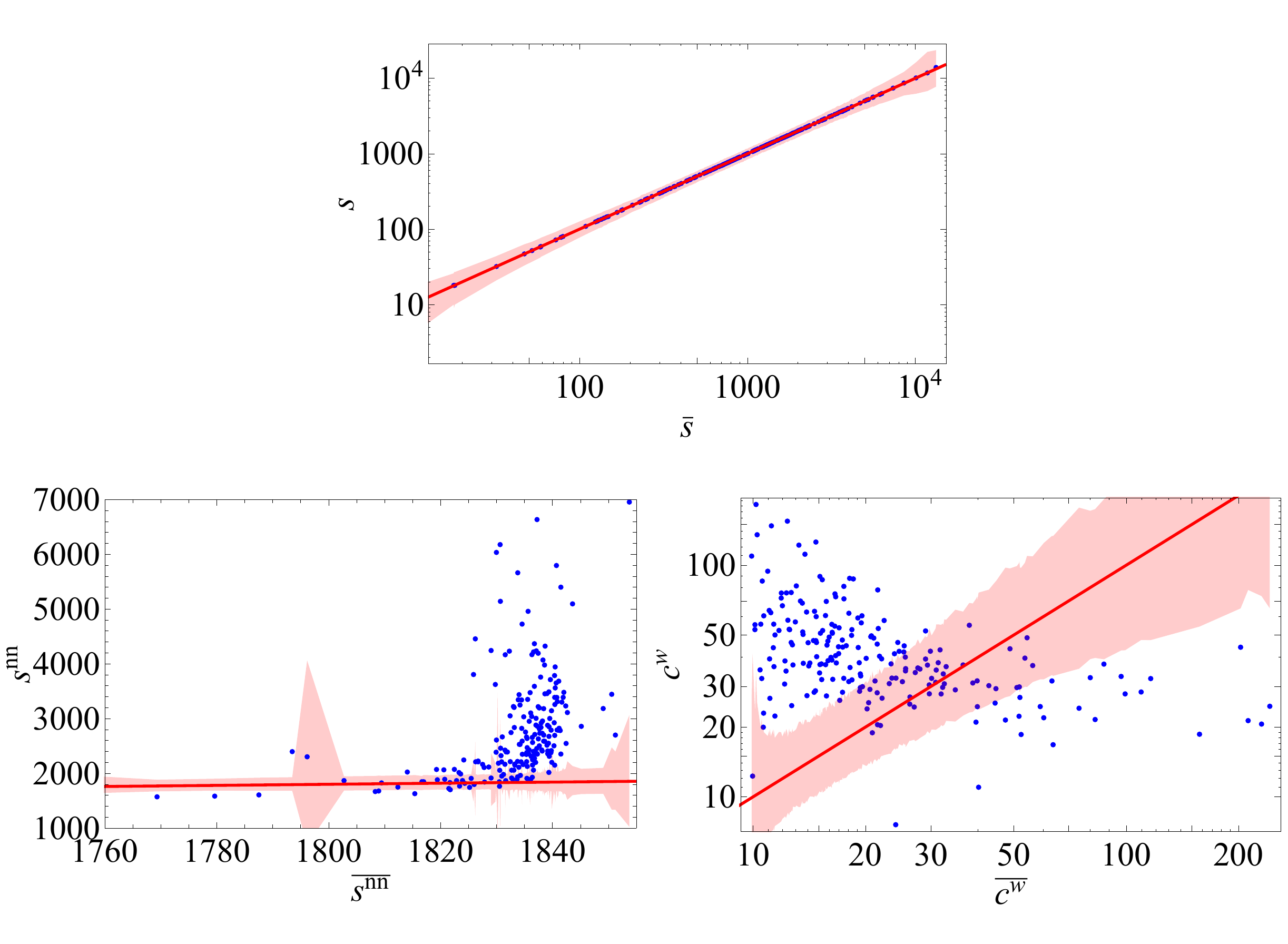}
\caption{Sampling weighted undirected networks with given strength sequence (Undirected Weighted Configuration Model). The example shown is the weighted network of liquidity reserves exchanges between Italian banks in 1999 \cite{Bankdata} (N = 215). The three panels show, for each node in the network, the comparison between the observed value and the sample average of the (constrained) strength (top), the (unconstrained) ANNS (bottom left) and the (unconstrained) weighted clustering coefficient (bottom right), for 1000 sampled networks. The 95\% confidence intervals of the distribution of the sampled quantities is shown in pink for each node.}
\label{wun_top2}
\end{figure*}

In fig. \ref{wun_w2} we show an application of this method to the same interbank network considered previously in figs. \ref{bun_a2} and \ref{bun_top2}, but now using its weighted representation \cite{Bankdata}. In this case we plot, for increasing numbers of sampled networks, the convergence of the sample average $\overline{w_{ij}}$ of each edge weight to its exact canonical expectation $\langle w_{ij}\rangle$. 
As for the example considered for the UBCM, generating 1000 matrices (bottom right) turns out to be enough to obtain a high level of accuracy for this network. 
This important check is impossible in microcanonical approaches, where there is no knowledge of the exact value of the expected weights.

Here as well, the average of the quantities of interest over the sample can be compared with the observed values.
As a preliminary check, the top plot of fig. \ref{wun_top2} confirms that, for the sample of 1000 matrices, the sample average of the strength of each node coincides with its observed value, and the confidence intervals are very narrow around the identity line. Thus the enforced constraints are correctly reproduced.
We can then properly use the UWCM as a null model to detect higher-order patterns in the network. 

In the bottom panels of fig. \ref{wun_top2} we show the average nearest neighbor strength (ANNS), $s_i^{nn}=\frac{\sum_ja_{ij}s_j}{k_i}$, and the weighted clustering coefficient, $c_i^w=\frac{\sum_{j,\:k}w_{ij}w_{jk}w_{ki}}{\sum_{j\neq k}w_{ij}w_{ik}}$.
In this case, in line with previous analyses of different networks \cite{mymethod,myPRE1,myPRE2,mynull,myenhanced}, we find that the UWCM is \emph{not} as effective as its binary counterpart in reproducing the observed higher-order properties, as clear from the presence of many outliers in the plots.
Since our previous checks ensure that the implementation of the null model is correct, we can safely conclude that the divergence between the null model and the real network is not due to an insufficient or incorrect sampling of the ensemble. 
Rather, it is a genuine signature of the fact that, in this network, the strength sequence alone is not enough in order to replicate higher-order quantities. So the strength sequence turns out to be less informative (about the whole weighted network) than the degree sequence is (about the binary projection of the same network). 

\begin{figure}[t]
\includegraphics[width=0.48\textwidth]{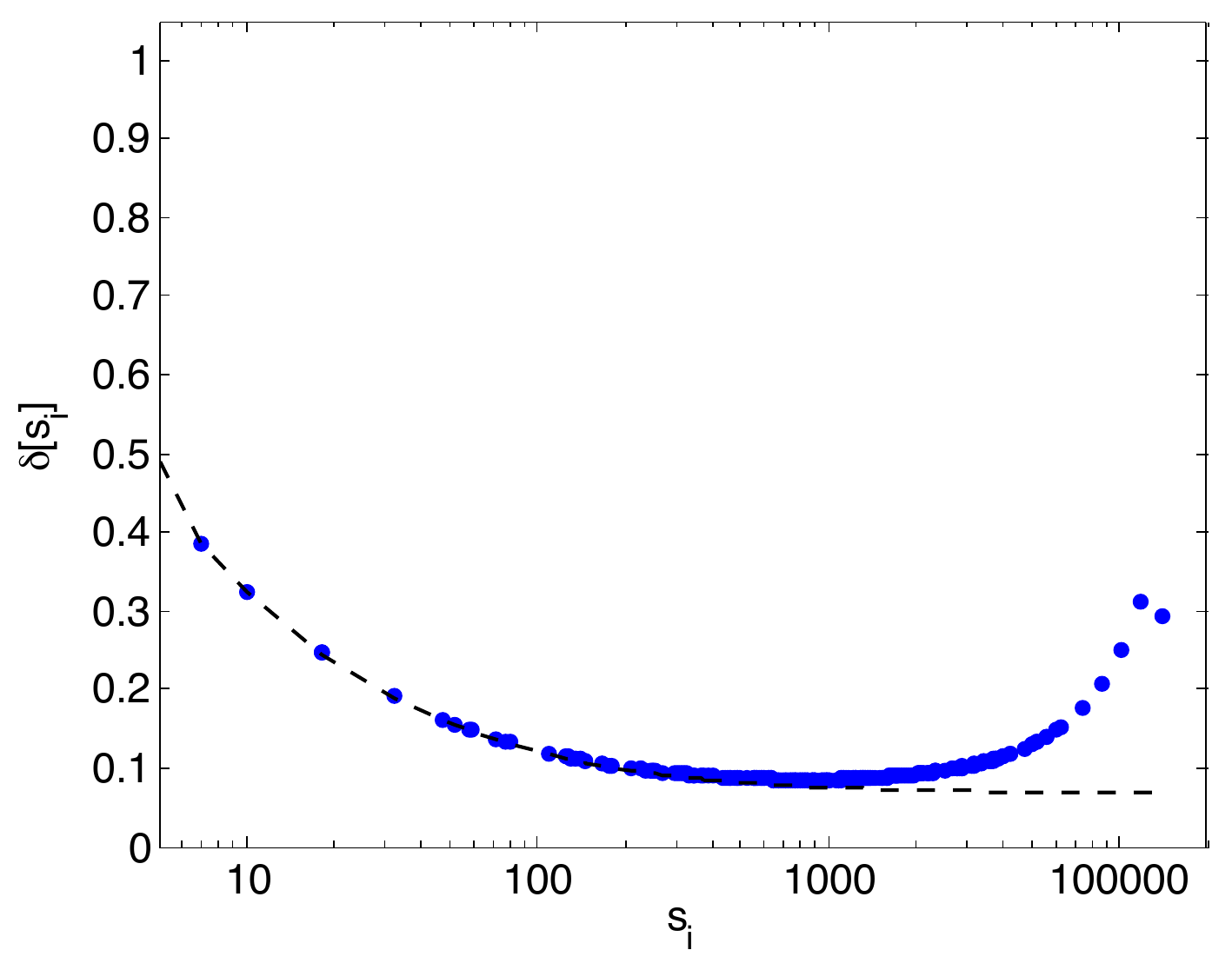}
\caption{Coefficient of variation $\delta[s_i]$ as a function of the strength $s_i$ for each node of the binary network of liquidity reserves exchanges between Italian banks in 1999 \cite{Bankdata} ($N$ = 215). 
The blue points are the exact values in eq.(\ref{eq:fluUWCM}), while the dashed curve is the lower bound in eq.(\ref{eq:boundUWCM}). The upper bound exceeds 1 and extends beyond the region shown.}
\label{fig:fluUWCM}
\end{figure}

We now come to the analysis of the canonical fluctuations. 
The ensemble variance of each strength $s_i$ is defined as $\sigma^2[s_i]\equiv\langle s_i^2\rangle-\langle s_i\rangle^2$, and the independence of pairs of nodes implies
\begin{eqnarray}
\sigma^2[s_i]&=&\sum_{j\ne i}\sigma^2[w_{ij}]=\sum_{j\ne i}(\langle w^2_{ij}\rangle-\langle w_{ij}\rangle^2)\nonumber\\
&=&\sum_{j \ne i} \frac{p_{ij}}{(1-p_{ij})^2}=\sum_{j\ne i}\langle w_{ij}\rangle(1+\langle w_{ij}\rangle)\nonumber\\
&=&s_i+\sum_{j\ne i}\langle w_{ij}\rangle^2.
\label{eq:varUWCM}
\end{eqnarray}
Therefore the relative fluctuations take the form
\begin{equation}
\delta[s_i]\equiv\frac{\sigma[s_i]}{s_i}=\sqrt{\frac{1}{s_i}+\frac{\sum_{j\ne i}\langle w_{ij}\rangle^2}{(\sum_{j\ne i}\langle w_{ij}\rangle)^2}}
\label{eq:fluUWCM}
\end{equation}
for $s_i>0$. A plot of $\delta[s_i]$ as a function of $s_i$ for the interbank network is shown in fig. \ref{fig:fluUWCM}. Unlike in the UBCM, here the relative fluctuations are found to be smaller for intermediate values of the strength.

When comparing eq.(\ref{eq:fluUWCM}) with eq.(\ref{eq:fluUBCM}), it is interesting to notice that the term ${\sum_{j\ne i}\langle w_{ij}\rangle^2}/{(\sum_{j\ne i}\langle w_{ij}\rangle)^2}$, while still being a participation ratio\footnote{In this case, the participatio ratio measures the inverse of the effective number of equally important terms in the sum ${\sum_{j\ne i}\langle w_{ij}\rangle}$. It equals $1$ if and only if there is only one nonzero term (complete concentration, which still implies $\langle k_i\rangle=1$ but \emph{not} $s_i=1$), while it equals $(N-1)^{-1}$ if and only if there are $N-1$ identical terms (complete homogeneity), i.e. $\langle w_{ij}\rangle=s_i/(N-1)$ for all $j\ne i$.}, is now predeced by a \emph{positive} sign. 
This implies that the bounds for $\delta[s_i]$ are quite different from those for $\delta[k_i]$ shown in eq.(\ref{eq:boundUBCM}):
\begin{equation}
\sqrt{\frac{1}{s_i}+\frac{1}{N-1}}\le\delta[s_i]\le\sqrt{\frac{1}{s_i}+1}.
\label{eq:boundUWCM}
\end{equation}
The allowed region for $\delta[s_i]$ is the one above the dashed line in  fig.\ref{fig:fluUWCM}, and extends beyond 1. We now find that the realized trend is very close to the \emph{lower} bound for small and intermediate values of the strength (again suggesting that in this regime our maximum-entropy method produces almost maximally homogeneous terms in the sum ${\sum_{j\ne i}\langle w_{ij}\rangle}$), while it exceeds the lower bound significantly for large values of the strength. 
In any case, since eq.(\ref{eq:boundUWCM}) implies that $\delta[s_i]$ cannot vanish for any value of $s_i$, we find evidence of the fact that for this model the microcanonical and canonical ensembles are \emph{always} not equivalent.  

\subsection{Weighted directed networks with given in-strength and out-strength sequences\label{sec:DWCM}}
We now consider weighted directed networks (WDNs), defined by a weight matrix $\mathbf{W}$ which is in general not symmetric.
Each node is now characterized by two strengths, the out-strength $s_i^{out}\equiv \sum_j w_{ij}$ and the in-strength $s_i^{in}\equiv \sum_j w_{ji}$. 
The \emph{Directed Weighted Configuration Model} (DWCM), the directed version of the UWCM, enforces the out- and in-strength sequences, $\{s^{out}_i\}_{i=1}^N$ and $\{s^{in}_i\}_{i=1}^N$, of a real-world network $\mathbf{W}^*$ \cite{mymethod,myPRE2,mynull}.
The model is widely used to detect modules and communities in real WDNs \cite{Newman 2010}.

In its canonical version, the DWCM is still characterized by eq.\eqref{eq:Pw} where ``$j<i$'' is replaced by ``$j\neq i$'' and now $p_{ij}\equiv x_iy_j$. 
The $2N$ unknown parameters $\mathbf{x}$ and $\mathbf{y}$ can be fixed by either maximizing the log-likelihood function \cite{mymethod}
\begin{eqnarray} 
\lambda(\mathbf{x},\mathbf{y})&\equiv&
\ln P(\mathbf{W}^*|\mathbf{x},\mathbf{y})=\label{DWCM}\\
&=&\sum_i \big[s_i^{out}  (\mathbf{W}^*) \ln x_i + s_i^{in}(\mathbf{W}^*)\ln y_i\big]\nonumber \\
&+&\sum_i\sum_{j \neq i} \ln (1-x_iy_j)\nonumber
\end{eqnarray} 
or solving the the $2N$ equations \cite{mymethod}
\begin{eqnarray} 
\langle s_i^{out} \rangle&=&\sum_{j \neq  i} \dfrac{x_iy_j}{1-x_iy_j}=s_i^{out}(\mathbf{W}^*)\quad \forall i \\
\langle s_i^{in} \rangle&=&\sum_{j \neq i} \dfrac{x_jy_i}{1-x_jy_i}=s_i^{in}(\mathbf{W}^*)\quad \forall i,
\end{eqnarray}
where in both cases the parameters $\mathbf{x}$ and $\mathbf{y}$ vary in the region defined by $0\le x_i y_j<1$ for all $i,j$ \cite{mywreciprocity}.

Once the unknown variables are found, we can implement an efficient and unbiased sampling scheme in the same way as for the UWCM, but now running over each pair of vertices \emph{twice} (i.e. in both directions). One can establish the weight of a link from vertex $i$ to vertex $j$ using the geometric distribution $p_{ij}^{w}(1-p_{ij})$, and the weight of the reverse link from $j$ to $i$ using the geometric distribution $p_{ji}^{w}(1-p_{ji})$, these two events being independent. 
Alternatively, as for the undirected case, one can construct these random events as a combination of fundamental Bernoulli trials with success probability $p_{ij}$ and $p_{ji}$.
Since this directed generalization of the undirected case is straightforward, we do not consider any explicit application.
However, we have explicitly included the DWCM model in the code (see Appendix).

We now come to the canonical fluctuations. In analogy with eq.(\ref{eq:varUWCM}), it is easy to show that the variances of $s^{out}_i$ and $s^{in}_i$ are given by 
\begin{eqnarray}
\sigma^2[s^{out}_{i}]&=& 
\sum_{j\ne i}\langle w_{ij}\rangle(1+\langle w_{ij}\rangle)=s^{out}_i+\sum_{j \ne i} \langle w_{ij}\rangle^2,\label{eq:varDWCM1}\\
\sigma^2[s^{in}_{i}]&=& \sum_{j\ne i}\langle w_{ji}\rangle(1+\langle w_{ji}\rangle)=s^{in}_i+\sum_{j \ne i} \langle w_{ji}\rangle^2.\label{eq:varDWCM2}
\end{eqnarray}
For $s^{out}_i>0$ and $s^{in}_i>0$, the relative fluctuations are
\begin{eqnarray}
\delta[s^{out}_i]&\equiv&\frac{\sigma[s^{out}_i]}{s^{out}_i}=\sqrt{\frac{1}{s^{out}_i}+\frac{\sum_{j\ne i}\langle w_{ij}\rangle^2}{(\sum_{j\ne i}\langle w_{ij}\rangle)^2}},\label{eq:fluDWCM1}\\
\delta[s^{in}_i]&\equiv&\frac{\sigma[s^{in}_i]}{s^{in}_i}=\sqrt{\frac{1}{s^{in}_i}+\frac{\sum_{j\ne i}\langle w_{ji}\rangle^2}{(\sum_{j\ne i}\langle w_{ji}\rangle)^2}}.\label{eq:fluDWCM2}
\end{eqnarray}
For the bounds of the above quantities, expressions similar to eq.(\ref{eq:boundUWCM}) apply, suggesting that the microcanonical and canonical versions of this ensemble are also not equivalent.

\subsection{Weighted directed networks with given strength sequences and reciprocity structure\label{sec:RWCM}}
In analogy with the binary case, we now consider the \emph{Reciprocal Weighted Configuration Model} (RWCM), which is a recently proposed null model that for the first time allows one to constrain the reciprocity structure in weighted directed networks \cite{mywreciprocity}. 
The RWCM enforces three strengths for each node: the non-reciprocated incoming strength, $s_i^{\leftarrow}\equiv\sum_j w_{ij}^\leftarrow$, the non-reciprocated outgoing strength, $s_i^{\rightarrow}\equiv\sum_j w_{ij}^\rightarrow$, and the reciprocated strength, $s_i^{\leftrightarrow}\equiv\sum_j w_{ij}^\leftrightarrow$ \cite{mywreciprocity}.
Such quantities are defined by means of three pair-specific variables: $w_{ij}^{\leftrightarrow}\equiv\min[w_{ij},\:w_{ji}]$ (reciprocated weight), $w_{ij}^{\rightarrow}\equiv w_{ij}-w_{ij}^{\leftrightarrow}$ and $w_{ij}^{\leftarrow}\equiv w_{ji}-w_{ij}^{\leftrightarrow}$ (non-reciprocated weights).

Despite its complexity, the RWCM is analytically solvable \cite{mywreciprocity} and the graph probability factorizes as:
\begin{equation}
P(\mathbf{W}|\mathbf{x},\mathbf{y},\mathbf{z})=\prod_i\prod_{j<i}\left[\frac{(x_iy_j)^{w_{ij}^{\rightarrow}}(x_jy_i)^{w_{ij}^{\leftarrow}}(z_iz_j)^{w_{ij}^{\leftrightarrow}}}{Z_{ij}(x_i,x_j,y_i,y_j,z_i,z_j)}\right]
\label{eq:RWCM}
\end{equation}
where $Z_{ij}(x_i,x_j,y_i,y_j,z_i,z_j)\equiv\frac{(1-x_ix_jy_iy_j)}{(1-x_iy_j)(1-x_jy_i)(1-z_iz_j)}$ is the node-pair partition function. 
The $3N$ unknown parameters $\mathbf{x}$, $\mathbf{y}$ and $\mathbf{z}$ must be determined either by maximizing the log-likelihood function
\begin{eqnarray}  
\lambda(\mathbf{x},\mathbf{y},\mathbf{z})&\equiv&
\ln P(\mathbf{W}^*|\mathbf{x},\mathbf{y},\mathbf{z})=\\
&=&\sum_i \big[s_i^{\rightarrow} (\mathbf{W}^*) \ln x_i + s_i^{\leftarrow}(\mathbf{W}^*)\ln y_i\nonumber\\
&+&s_i^{\leftrightarrow}(\mathbf{W}^*) \ln z_i\big] \!-\! \sum_i\sum_{j<i}\ln Z_{ij}(x_i,x_j,y_i,y_j,z_i,z_j)\nonumber
\end{eqnarray}
or by solving the $3N$ equations: 
\begin{eqnarray}
\langle s_i^{\rightarrow} \rangle&=&\sum_{j \neq i} \frac{x_iy_j(1-x_jy_i)}{(1-x_iy_j)(1-x_ix_jy_iy_j)}=s_i^{\rightarrow}(\mathbf{W}^*)\quad\forall i \qquad\\
\langle s_i^{\leftarrow} \rangle&=&\sum_{j \neq i} \frac{x_jy_i(1-x_iy_j)}{(1-x_jy_i)(1-x_ix_jy_iy_j)}=s_i^{\leftarrow}(\mathbf{W}^*)\quad\forall i \\
\langle s_i^{\leftrightarrow} \rangle&=&\sum_{j \neq i} \frac{z_iz_j}{1-z_iz_j}=s_i^{\leftrightarrow}(\mathbf{W}^*)\quad\forall i.
\end{eqnarray}
Here, the parameters $\mathbf{x}$,  $\mathbf{y}$ and $\mathbf{z}$ vary in the region defined by $0\le x_i y_j<1$ and $0\le z_i z_j<1$ for all $i,j$ \cite{mywreciprocity}.

Equation \eqref{eq:RWCM} shows that pairs of nodes are independent, and that the probability that the nodes $i$ and $j$ are connected via a combination of weighted edges of the form $(w_{ij}^\leftarrow,w_{ij}^\rightarrow,w_{ij}^\leftrightarrow)$
is $\left[\frac{(x_iy_j)^{w_{ij}^{\rightarrow}}(x_jy_i)^{w_{ij}^{\leftarrow}}(z_iz_j)^{w_{ij}^{\leftrightarrow}}}{Z_{ij}(x_i,x_j,y_i,y_j,z_i,z_j)}\right]$ (where, as usual, all the parameters are intended to be the ones maximizing the likelihood).
Also, note that $w_{ij}^{\leftarrow}$ and $w_{ij}^{\rightarrow}$ cannot be both nonzero, but they are independent of $w_{ij}^{\leftrightarrow}$ (the joint distribution of these three quantities shown above is not simply a multivariate geometric distribution).

The above observations allow us to define an unbiased sampling scheme, even if more complicated than the ones described so far.
For each pair of nodes $i,j$, we define a procedure in three steps. 
First, we draw the reciprocal weight $w_{ij}^{\leftrightarrow}$ from the geometric distribution  $(z_iz_j)^{w_{ij}^{\leftrightarrow}}(1-z_iz_j)$ (or equivalently, from the composition of Bernoulli distributions as discussed for the UWCM).
Second, we focus on the \emph{mere existence} of non-reciprocated weights (irrespective of their magnitude). We randomly select one of these three (mutually excluding) events: we establish the absence of any non-reciprocated weight between $i$ and $j$ ($w_{ij}^{\rightarrow}=0$, $w_{ij}^{\leftarrow}=0$) with probability $\frac{(1-x_iy_j)(1-x_jy_i)}{1-x_ix_jy_iy_j}$, we establish the existence of a non-reciprocated weight from $i$ to $j$ ($w_{ij}^{\rightarrow}>0$, $w_{ij}^{\leftarrow}=0$) with probability $\frac{x_iy_j(1-x_jy_i)}{1-x_ix_jy_iy_j}$, we establish the existence of a non-reciprocated weight from $j$ to $i$ ($w_{ij}^{\rightarrow}=0$, $w_{ij}^{\leftarrow}>0$) with probability $\frac{x_jy_i(1-x_iy_j)}{1-x_ix_jy_iy_j}$.
Third, if a non-reciprocated connection has been established (i.e. if its weight $w$ is positive) we then focus on the value to be assigned to it (i.e. on the extra weight $w-1$). 
If $w_{ij}^{\rightarrow}>0$, we draw the weight $w_{ij}^{\rightarrow}$ from a geometric distribution  $(x_iy_j)^{w_{ij}^{\rightarrow}-1}(1-x_iy_j)$ (shifted to strictly positive integer values of $w_{ij}^{\rightarrow}$ via the rescaled exponent), while if $w_{ij}^{\leftarrow}>0$ we draw the weight $w_{ij}^{\leftarrow}$ from the distribution $(x_jy_i)^{w_{ij}^{\leftarrow}-1}(1-x_jy_i)$. 

The recipe described above is still of complexity $O(N^2)$ and allows us to sample the canonical ensemble of the RWCM in an unbiased and efficient way. It should be noted that the microcanonical analogue of this algorithm has not been proposed so far.
As for the DWCM, we show no explicit application, even if the entire algorithm is available in our code (see Appendix).

In this model, the canonical fluctuations are somewhat more compicated than in the previous models. The variances of the constraints are
\begin{eqnarray}
\sigma^2[s_i^\rightarrow] &=& \sum_{j \ne i} \frac{x_iy_j(1-x_jy_i)(1-x_i^2x_jy_iy_j^2)}{(1-x_iy_j)^2(1-x_ix_jy_iy_j)^2},\\
\sigma^2[s_i^\leftarrow] &=& \sum_{j \ne i} \frac{x_jy_i(1-x_iy_j)(1-x_ix_j^2y_i^2y_j)}{(1-x_jy_i)^2(1-x_ix_jy_iy_j)^2}\\
\sigma^2[s_i^\leftrightarrow]& =& \sum_{j \ne i} \frac{z_{i}z_{j}}{(1-z_{i}z_{j})^2}.
\end{eqnarray}
While for the variance of the reciprocated weight we can still write $\sigma^2[w^\leftrightarrow_{ij}]=\langle w^\leftrightarrow_{ij}\rangle(1+\langle w^\leftrightarrow_{ij}\rangle)$ in analogy with the UWCM and DWCM, similar relations do not hold for the non-reciprocated weights. 
However, since $x_i y_j<1$ for all $i,j$, it is easy to show that $\sigma^2[w^\rightarrow_{ij}]>\langle w^\rightarrow_{ij}\rangle(1+\langle w^\rightarrow_{ij}\rangle)$ and $\sigma^2[w^\leftarrow_{ij}]>\langle w^\leftarrow_{ij}\rangle(1+\langle w^\leftarrow_{ij}\rangle)$.
This still allows us to obtain a lower bound for all quantities as in the other weighted models, by using
\begin{eqnarray}
\sigma^2[s^{\to}_{i}]&>& \sum_{j \neq i} \langle w^{\rightarrow}_{ij}\rangle(1+\langle w^{\rightarrow}_{ij}\rangle)=s^{\rightarrow}_i+\sum_{j \neq i} \langle w^{\rightarrow}_{ij}\rangle^2,\label{eq:varRWCM1}\\
\sigma^2[s^{\gets}_{i}]&>& \sum_{j \neq i} \langle w^{\leftarrow}_{ij}\rangle(1+\langle w^{\leftarrow}_{ij}\rangle)=s^{\leftarrow}_i+\sum_{j \neq i} \langle w^{\leftarrow}_{ij}\rangle^2,\label{eq:varRWCM2}\\
\sigma^2[s^{\leftrightarrow}_{i}]&=& \sum_{j \neq i} \langle w^{\leftrightarrow}_{ij}\rangle(1+\langle w^{\leftrightarrow}_{ij}\rangle)=s^{\leftrightarrow}_i+\sum_{j \neq i} \langle w^{\leftrightarrow}_{ij}\rangle^2.
\label{eq:varRWCM3}
\end{eqnarray}
Then, for $s^{\to}_i>0$, $s^{\gets}_i>0$ and $s^{\leftrightarrow}_i>0$, we get
\begin{eqnarray}
\delta[s^{\rightarrow}_i]&\equiv&\frac{\sigma[s^{\rightarrow}_i]}{s^{\rightarrow}_i}>\sqrt{\frac{1}{s^{\rightarrow}_i}+\frac{\sum_{j\ne i}\langle w^{\rightarrow}_{ij}\rangle^2}{(\sum_{j\ne i}\langle w^{\rightarrow}_{ij}\rangle)^2}},\label{eq:fluRWCM1}\\
\delta[s^{\leftarrow}_i]&\equiv&\frac{\sigma[s^{\leftarrow}_i]}{s^{\leftarrow}_i}>\sqrt{\frac{1}{s^{\leftarrow}_i}+\frac{\sum_{j\ne i}\langle w^{\leftarrow}_{ij}\rangle^2}{(\sum_{j\ne i}\langle w^{\leftarrow}_{ij}\rangle)^2}},\label{eq:fluRWCM2}\\
\delta[s^{\leftrightarrow}_i]&\equiv&\frac{\sigma[s^{\leftrightarrow}_i]}{s^{\leftrightarrow}_i}=\sqrt{\frac{1}{s^{\leftrightarrow}_i}+\frac{\sum_{j\ne i}\langle w^{\leftrightarrow}_{ij}\rangle^2}{(\sum_{j\ne i}\langle w^{\leftrightarrow}_{ij}\rangle)^2}}.\label{eq:fluRWCM3}
\end{eqnarray}
Therefore, for all three quantities, a lower bound of the form $\delta[s_i]\ge \sqrt{\frac{1}{s_i}+\frac{1}{N-1}}$ still applies, as in eq.(\ref{eq:boundUWCM}). 
This suggests that, for this model as well, the microcanonical and canonical ensembles are not equivalent.

\subsection{Weighted undirected networks with given strengths and degrees\label{sec:UECM}}
We finally consider a `mixed' null model of weighted networks with both binary (degree sequence $\{k_i\}_{i=1}^N$) and weighted (strength sequence $\{s_i\}_{i=1}^N$) constraints. We only consider undirected networks for simplicity, but the extension to the directed case is straightforward.
The ensemble of weighted undirected networks with given strengths and degrees has been recently introduced as the \emph{(Undirected) Enhanced Configuration Model} (UECM) \cite{myenhanced,myextensive}.

This model, which is based on analytical results derived in \cite{mybosefermi}, is of great importance for the problem of \emph{network reconstruction} from partial node-specific information \cite{myenhanced}.
As we have also illustrated in fig.\ref{wun_top2}, the knowledge of the strength sequence alone is in general not enough in order to reproduce the higher-order properties of a real-world weighted network \cite{myPRE2,mynull}. Usually, this is due to the fact that the expected topology is much denser than the observed one (often the expected network is almost fully connected).
By contrast, it turns out that the simultaneous specification of strengths and degrees, by constraining the local connectivity to be  consistent with the observed one, allows a dramatically improved reconstruction of the higher-order structure of the original weighted network \cite{myenhanced,myextensive}.

\begin{figure*}[t!]
\includegraphics[width=0.8\textwidth]{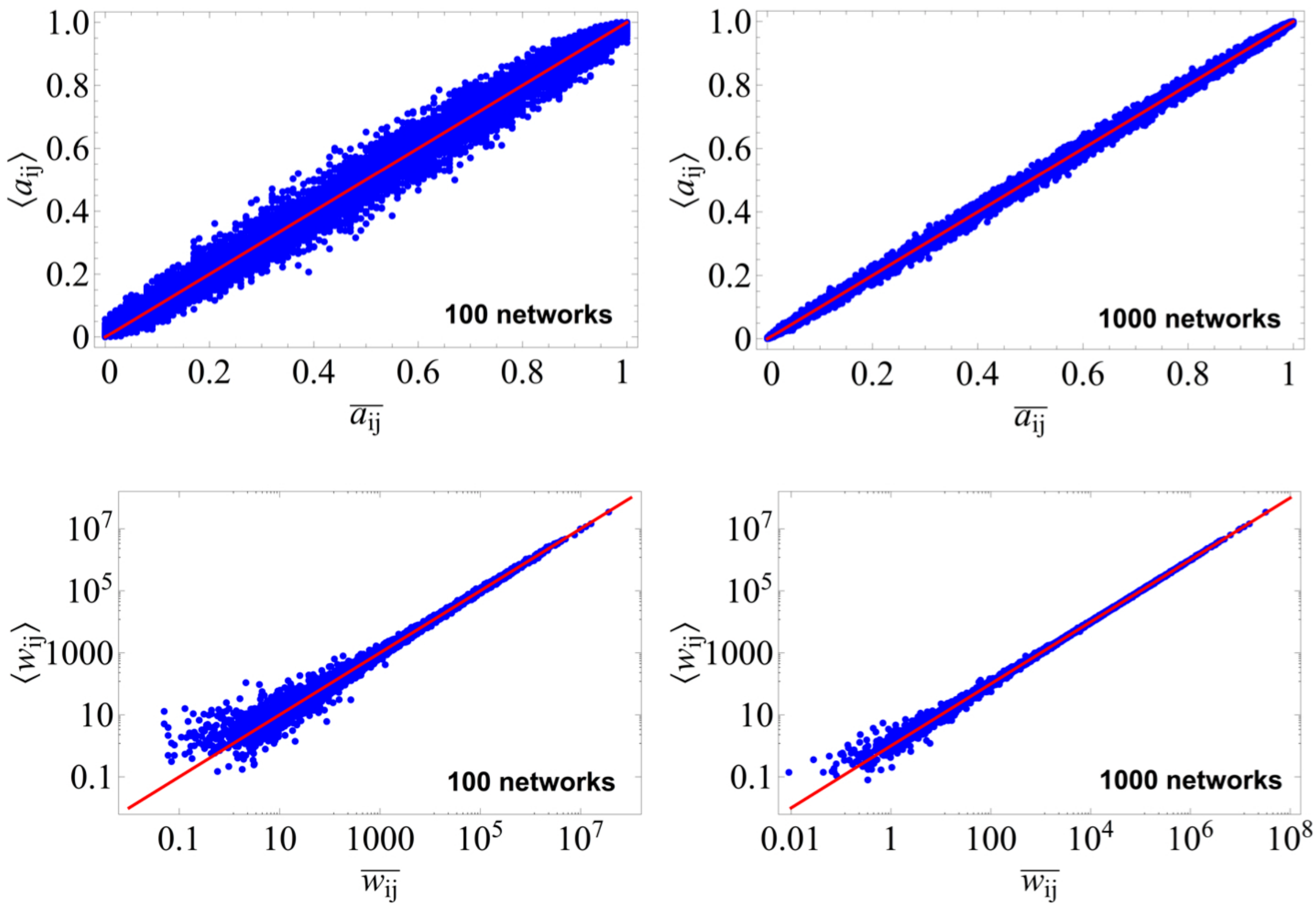}
\caption{Sampling weighted undirected networks with given degree and strength sequences (Undirected Enhanced Configuration Model). The example shown is the weighted World Trade Web (N = 162) \cite{wtwd}. The top panels show the convergence of the sample average $\overline{a_{ij}}$ of each entry of the adjacency matrix to its exact canonical expectation $\langle p_{ij}\rangle$, for 100 (left) and 1000 (right) sampled matrices. 
The bottom panels show the convergence of the sample average $\overline{w_{ij}}$ of each entry of the weight matrix to its exact canonical expectation $\langle w_{ij}\rangle$, for 100 (left) and 1000 (right) sampled networks. The identity line is shown in red.}
\label{mix_aw2}
\end{figure*}
\begin{figure*}[t!]
\includegraphics[width=0.8\textwidth]{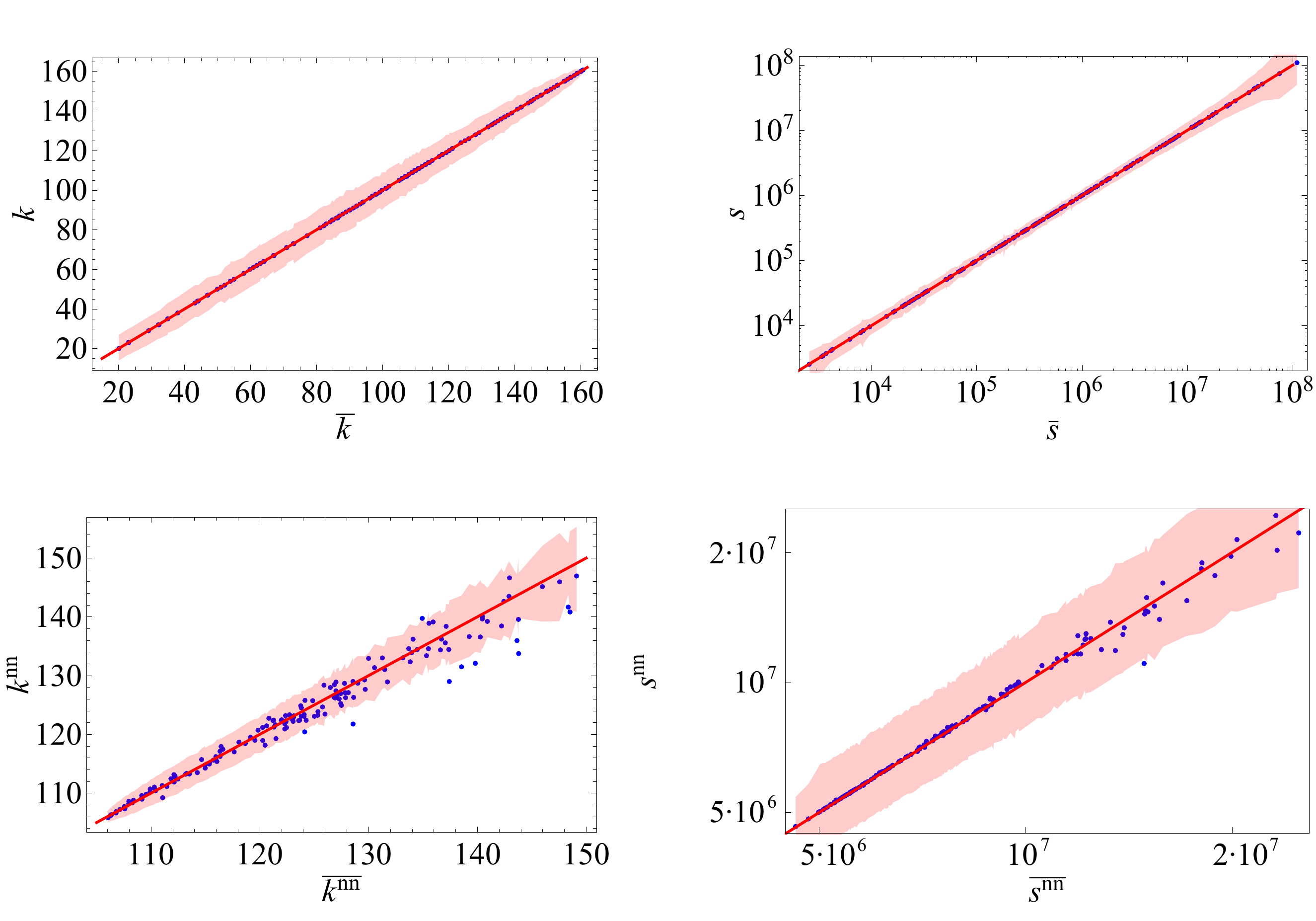}
\caption{Sampling weighted undirected networks with given degree and strength sequences (Undirected Enhanced Configuration Model). The example shown is the weighted World Trade Web (N = 162) \cite{wtwd}. 
The four panels show, for each node in the network, the comparison between the observed value and the sample average of the (constrained) degree (top left), the (constrained) strength (top right), the (unconstrained) ANND (bottom left) and the (unconstrained) ANNS (bottom right), for 1000 sampled networks. The 95\% confidence intervals of the distribution of the sampled quantities is shown in pink for each node.}
\label{mix_top2}
\end{figure*}

This very promising result calls for an efficient implementation of the UECM. 
We now describe an appropriate sampling procedure.
The probability distribution characterizing the UECM is halfway between a Bernoulli (Fermi-like) and a geometric (Bose-like) distribution  \cite{mybosefermi}, and reads
\begin{equation}
P(\mathbf{W}|\mathbf{x},\mathbf{y})\!=\!\prod_i\prod_{j<i}\left[\frac{(x_{i}x_{j})^{\Theta(w_{ij})}(y_{i}y_{j})^{w_{ij}}(1-y_{i}y_{j})}{1-y_{i}y_{j}+x_{i}x_{j}y_{i}y_{j}}\right].
\label{eq_mix}
\end{equation}
As usual, the $2N$ unknown parameters must be determined either by maximizing the log-likelihood function 
\begin{eqnarray} 
\lambda(\mathbf{x},\mathbf{y})&\equiv&\ln P(\mathbf{W}^*|\mathbf{x},\mathbf{y})=\nonumber\\
&=&\sum_i [k_i(\mathbf{W}^*) \ln x_i + s_i(\mathbf{W}^*)\ln y_i]\nonumber\\
&+&\sum_{i}\sum_{j<i} \ln \frac{1-y_iy_j}{(1-y_iy_j + x_ix_jy_iy_j)}
\end{eqnarray}
or by solving the $2N$ equations  \cite{myenhanced}:
\begin{eqnarray}
\langle k_i \rangle&=&\sum_{j\neq i} p_{ij}=k_i(\mathbf{W}^*)\quad\forall i\label{smix}\\
\langle s_i\rangle &=&\sum_{j\neq i} \frac{p_{ij}}{1-y_iy_j}=s_i(\mathbf{W}^*)\quad\forall i
\label{smix2}
\end{eqnarray}
where $p_{ij}\equiv \frac{x_{i}x_{j}y_{i}y_{j}}{1-y_{i}y_{j}+x_{i}x_{j}y_{i}y_{j}}$.
Here, the parameters $\mathbf{x}$ and $\mathbf{y}$ vary in the region $x_i\ge 0$ for all $i$ and $0\le y_i y_j<1$ for all $i,j$ respectively \cite{myenhanced}.

In order to define an unbiased sampling scheme, we note that eq. \eqref{eq_mix} highlights the two key ingredients of the UECM, respectively controlling for the probability that a link of any weight exists and, if so, that a specific positive weight is there. 
The probability to generate a link of weight $w$ between the nodes $i$ and $j$ is
\begin{equation}
q_{ij}(w) = \left\{ \begin{array}{ll}
1-p_{ij} & \quad\textrm{if}\quad w=0\\
p_{ij}(y_{i}y_{j})^{w-1}(1-y_{i}y_{j}) &\quad\textrm{if}\quad w>0\end{array} \right.
\nonumber
\end{equation}
The above expression identifies two key steps: the model is equivalent to one where the `first link' (of unit weight) is extracted from a Bernoulli distribution with probability $p_{ij}$ and where the `extra weight' ($w_{ij}-1$) is extracted from a  geometric distribution (shifted to the strictly positive integers) with parameter $y_{i}y_{j}$. 
As all the other examples discussed so far, this algorithm can be easily implemented.

In fig. \ref{mix_aw2} we provide an application of this method to the World Trade Web \cite{myPRE1,myPRE2,wtwd}. We show the convergence of the sample averages ($\overline{a_{ij}}$ and $\overline{w_{ij}}$) of the entries of both binary and weighted adjacency matrices to their exact canonical expectations ($\langle a_{ij}\rangle$ and $\langle w_{ij}\rangle$ respectively). 
As in the previous cases, generating 1000 matrices is enough to guarantee a tight convergence of the sample averages to their exact values (in any case, this accuracy can be quantified and improved by sampling more matrices).

For this sample of 1000 matrices, in the top plots (two in this case) of fig. \ref{mix_top2} we confirm that both the binary and weighted constraints are well reproduced by the sample averages.
When we use this null model to check for higher-order patterns in this network, we find that two important topological quantities of interest (ANND and ANNS, bottom panels of fig. \ref{mix_top2}) are well replicated by the model.
These results are consistent with what is obtained analytically by using the same canonical null model on the same network \cite{myextensive}.
Moreover, in this case we can calculate confidence intervals besides expected values (for instance, in fig. \ref{mix_top2} we can clearly identify outliers that are otherwise undetected), and do this for any desired topological property, not only those whose expected value is analytically computable.
Our method therefore represents an improved algorithm for the unbiased reconstruction of weighted networks from strengths and degrees \cite{myenhanced}.

\begin{figure*}[t]
\includegraphics[width=0.48\textwidth]{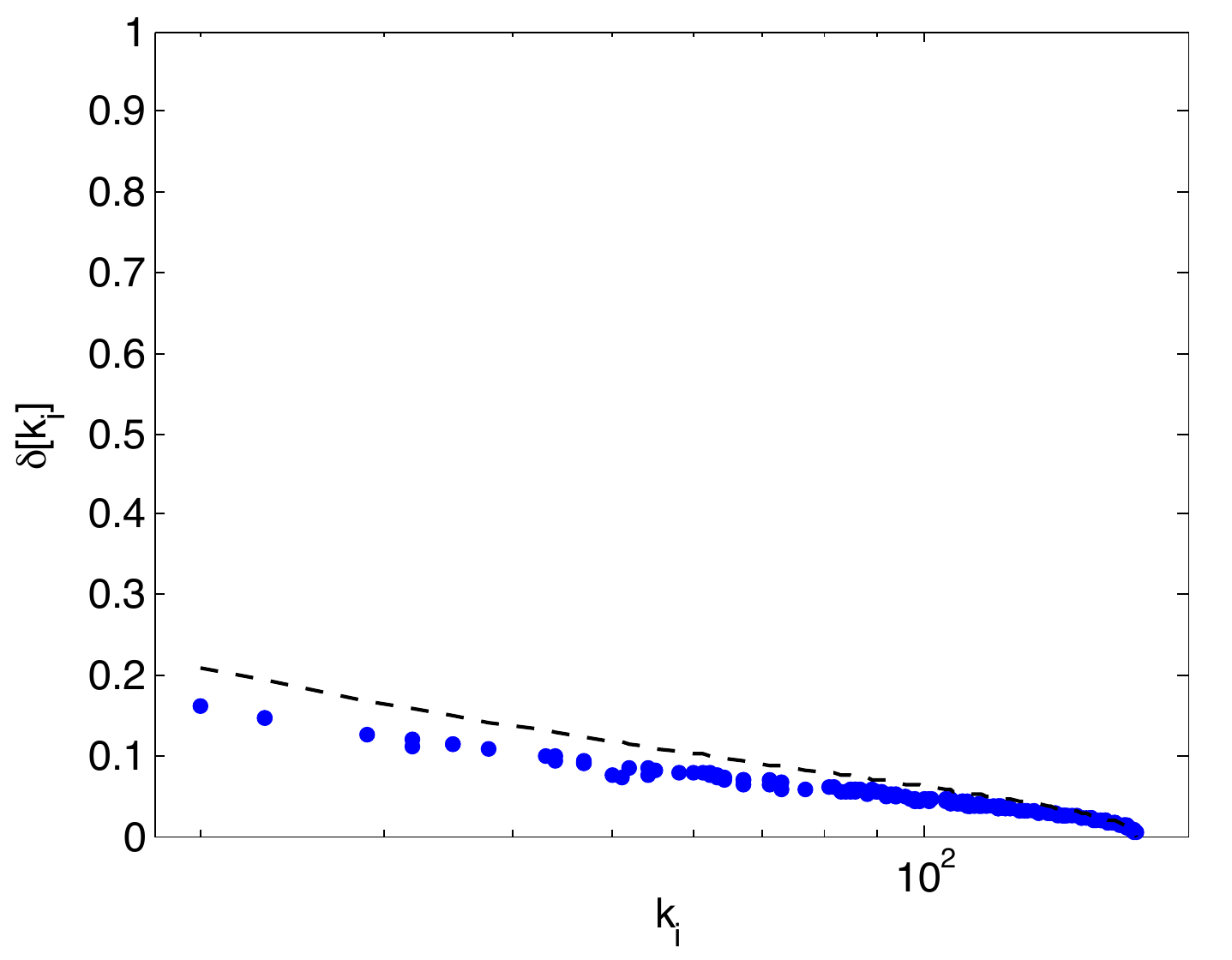}
\includegraphics[width=0.48\textwidth]{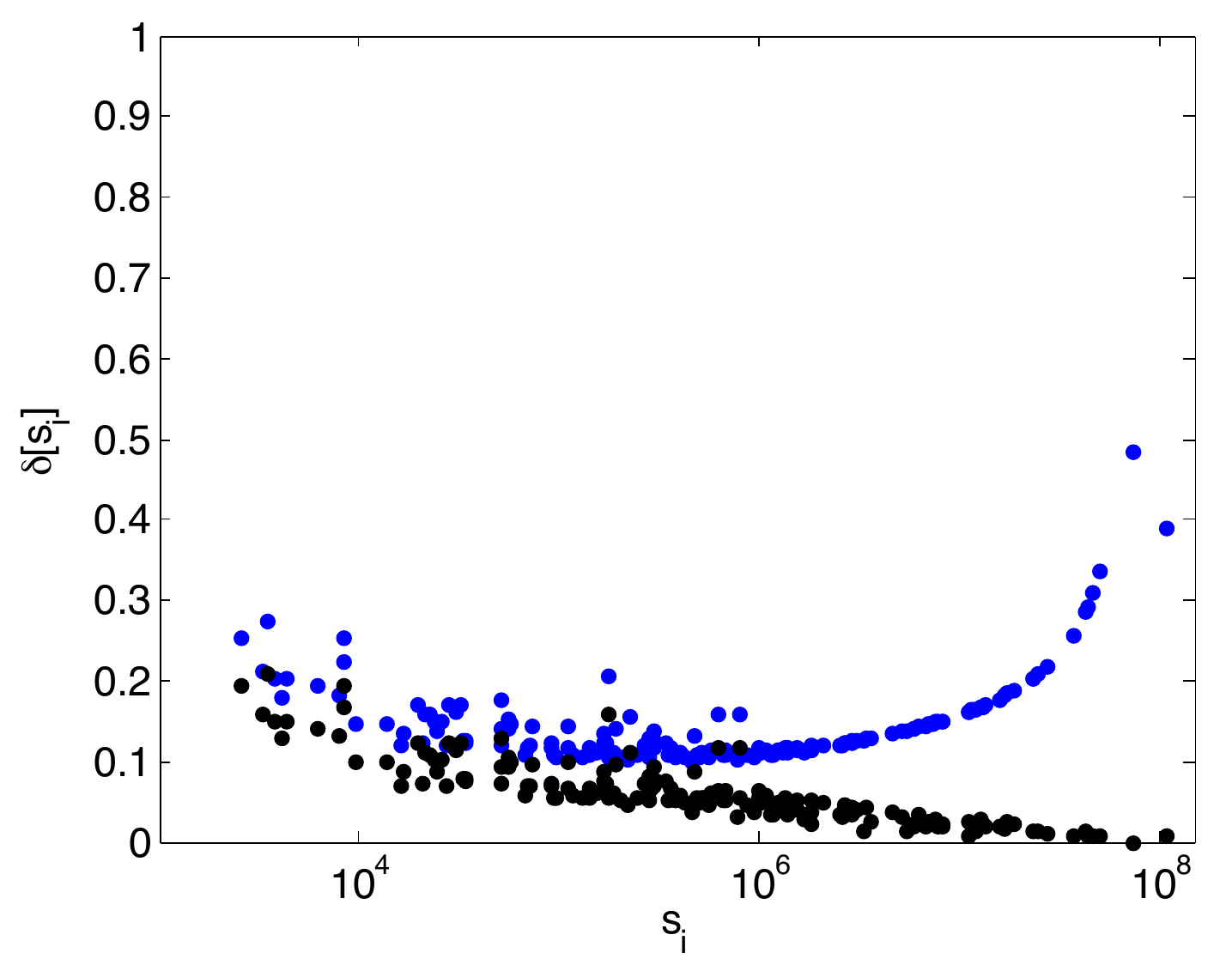}
\caption{Coefficients of variation $\delta[k_i]$  (left) and $\delta[s_i]$ (right), plotted as a function of the degree $k_i$ and the strength $s_i$ respectively, for each node of the weighted World Trade Web ($N = 162$) \cite{wtwd}. 
The blue points indicate the exact values, while the dashed curve (left) and black points (right) indicate in both cases the value $\sqrt{\frac{1}{k_i}-\frac{1}{N-1}}$, which is a strict upper bound for $\delta[k_i]$ and a reference value, typically exceeded, for $\delta[s_i]$.}
\label{fig:fluUECM}
\end{figure*}

The canonical fluctuations in this ensemble can be also calculated analytically. For the variance of the degrees, we can still exploit the expression $\sigma^2[a_{ij}]=p_{ij}(1-p_{ij})$.
For the variance of the strengths, we can use the definition $\sigma^2[w_{ij}]=\langle w_{ij}^2\rangle-\langle w_{ij}\rangle^2$, which however leads to a more complicated expression in this case. Using the relation $\langle w_{ij}\rangle=p_{ij}/(1-y_iy_j)$, the end result can be expressed as follows:
\begin{eqnarray}
\sigma^2[k_i]& =& \sum_{j \ne i} p_{ij}(1-p_{ij}),\\
\sigma^2[s_i]& =&\sum_{j\ne i}\frac{p_{ij}(1+y_{i}y_{j}-p_{ij})}{(1-y_{i}y_{j})^2}\nonumber\\
&=& \sum_{j \ne i} \langle w_{ij}\rangle\left(\frac{1+y_iy_j}{1-y_iy_j}-\langle{w_{ij}}\rangle\right).
\end{eqnarray}
Since $(1+y_iy_j)/(1-y_iy_j)\ge 1$, we can obtain the following relations for the relative fluctuations:
\begin{eqnarray}
\delta[k_i]&\equiv&\frac{\sigma[k_i]}{k_i}=\sqrt{\frac{1}{k_i}-\frac{\sum_{j\ne i}p_{ij}^2}{(\sum_{j\ne i}p_{ij})^2}},\label{eq:fluUECM1}\\
\delta[s_i]&\equiv&\frac{\sigma[s_i]}{s_i}\ge\sqrt{\frac{1}{s_i}-\frac{\sum_{j\ne i}\langle{w_{ij}}\rangle^2}{(\sum_{j\ne i}\langle{w_{ij}}\rangle)^2}}.\label{eq:fluUECM2}
\end{eqnarray}
So $\delta[k_i]$ retains the same expression valid for the UBCM and all the other ensembles of binary graphs considered previously, which in turn leads to the same bounds as in eq.(\ref{eq:boundUBCM}). This is confirmed in fig.\ref{fig:fluUECM}.
By contrast, $\delta[s_i]$ has a more complicated form, which differs from that valid for the UWCM and does not lead to simple expressions for the upper and lower bounds. Also note the presence of a \emph{minus} sign in eq.(\ref{eq:fluUECM2}).
What can be concluded relatively easily is that, in the ideal limit $y_i\to 0$ (corresponding to very small values of $s_i$), we have $\langle w_{ij}\rangle\to p_{ij}$ which implies $s_i\to k_i$ and $\delta[s_i]\to \delta[k_i]$. This means that, in this extreme (and typically unrealized) limit, $\delta[s_i]$ behaves as $\delta[k_i]$, so it has the same upper bound $\sqrt{\frac{1}{k_i}-\frac{1}{N-1}}$. 
However, since $y_i$ is typically larger than zero, this bound is systematically exceeded, especially for large values of $s_i$. 
This is also confirmed in fig.\ref{fig:fluUECM}.
As in the other models, the non-vanishing of the fluctuations suggests that the microcanonical and canonical ensembles are not equivalent.

\section{Microcanonical considerations}
In this section we come back to the difference between canonical and microcanonical approaches to the sampling of network ensembles and discuss how, at least in principle, our method can be turned into an unbiased microcanonical one.

We provided evidence that, for all the models considered in this paper, the canonical and microcanonical ensembles are \emph{not} equivalent (see also \cite{mybreaking} for a recent mathematical proof of nonequivalence for the UBCM).
This result implies that choosing between microcanonical and canonical approaches to the sampling of network ensembles is not only a matter of (computational) convenience, but also a theoretical issue that should be addressed more formally. 

To this end, we recall that microcanonical ensembles describe isolated systems that do not interact with an external `heat bath' or `reservoir'.
In ordinary statistical physics, this means that there is no exchange of energy with the external world.
In our setting, this means that microcanonical approaches do not contemplate the possibility that the network interacts with some external `source of error', i.e. that the value of the enforced constraints might be affected by errors or missing entries in the data.
When present, such errors (e.g. a missing link, implying a wrong value of the degree of two nodes) are propagated to the entire collection of randomized networks, with the result that the `correct' network is not included in the microcanonical collection of graphs on which inference is being made.

By contrast, besides being unbiased and mathematically tractable, our canonical approach is also the most appropriate choice if one wants to account for possible errors in the data, since canonical ensembles appropriately describe systems in contact with an external reservoir (source of errors) affecting the value of the constraints. 
While in presence of even small errors microcanonical methods assign zero probability to the `uncorrupted' configuration and to all the configurations with the same value of the constraints, our method assigns these configurations a probability which is only slightly smaller than the (maximum) probability assigned to the set of configurations consistent with the observed (`corrupted') one.
These considerations suggest that, given its simplicity, elegance, and ability to deal with potential errors in the data, the use of the canonical ensemble should be preferred to that of the microcanonical one.

Nonetheless, it is important to note that, at least in principle, our canonical method can also be used to provide unbiased microcanonical expectations, if theoretical considerations suggest that the microcanonical ensemble is more appropriate in some specific cases. In fact, if the sampled configurations that do not satisfy the chosen constraints exactly are discarded, what remains is precisely an unbiased (uniform) sample of the microcanonical ensemble of networks defined by the same constraints (now enforced sharply). 
The sample is uniform because all the microcanonical configurations have the same probability of occurrence in the canonical ensemble (since all probabilities, as we have shown, depend only on the value of the realized constraints).
The same kind of analysis presented in this paper can then be repeated to obtain the microcanonical expectations. 
In the rest of this section, we discuss some advantages and limitations of this approach.

As a guiding principle, one should bear in mind that, to be feasible, a microcanonical sampling based on our method requires that the number $R_c$ of canonical realizations to be sampled (among which only a number $R_m<R_c$ of microcanonical ones will be selected) is not too large, especially because for each canonical realization one must (in the worst-case scenario) do $O(N)$ checks to ensure that each constraint matches the observed value exactly (the actual number is smaller, since all the checks after the first unsuccessful one can be aborted).

We first discuss the relation between $R_c$ and $R_m$.
Let $\mathbf{G}$ denote a generic graph (either binary or weighted) in the canonical ensemble, and $\mathbf{G}^*$ the observed network that needs to be randomized.
Let $\mathbf{h}$ formally denote a generic vector of chosen constraints, and let $\mathbf{h}^*\equiv \mathbf{h}(\mathbf{G}^*)$ indicate the observed values of such constraints.
Similarly, let $\mathbf{\theta}$ denote the generic vector of Lagrange multipliers (hidden variables) associated with $\mathbf{h}$, and let $\mathbf{\theta}^*$ indicate the vector of their likelihood-maximizing values enforcing the constraints $\mathbf{h}^*$.
On average, out of $R_c$ canonical realizations, we will be left with a number
\begin{equation}
R_m=Q(\mathbf{h}^*)R_c
\label{eq:rm}
\end{equation}
of microcanonical realizations, where 
$Q(\mathbf{h}^*)$ is the probability to pick a graph in the canonical ensemble that matches the constraints $\mathbf{h}^*$ exactly.
This probability reads
\begin{equation}
Q(\mathbf{h}^*)=\!\!\!
\sum_{\mathbf{G}/\mathbf{h}(\mathbf{G})=\mathbf{h}^*}
\!\!\!P(\mathbf{G}|\mathbf{\theta}^*)=N_m(\mathbf{h}^*)P(\mathbf{G}^*|\mathbf{\theta}^*)
\label{eq:qc}
\end{equation}
where $P(\mathbf{G}|\mathbf{\theta}^*)$ is the probability of graph $\mathbf{G}$ in the canonical ensemble, and $N_m(\mathbf{h}^*)$ is the number of microcanonical networks matching the constraints $\mathbf{h}^*$ exactly (i.e. the number of graphs with given $\mathbf{h}^*$).
Inserting eq. \eqref{eq:qc} into eq. \eqref{eq:rm} and inverting, we find that the value of $R_c$ required to distill $R_m$ microcanonical graphs is 
\begin{equation}
R_c=\frac{R_m}{N_m(\mathbf{h}^*)P(\mathbf{G}^*|\mathbf{\theta}^*)}
\end{equation}

Note that $P(\mathbf{G}^*|\mathbf{\theta}^*)$ is nothing but the maximized likelihood of the observed network, which is automatically measured in our method. This is typically an extremely small number: for the networks in our analysis, it ranges between $3.8\cdot 10^{-36468}$ (World Trade Web) and $4.9\cdot 10^{-3499}$ (binary interbank network). 
On the other hand, the number $N_m(\mathbf{h}^*)$ is very large (compensating the small value of the likelihood) but unknown in the general case: enumerating all graphs with given (sharp) properties is an open problem in combinatorics, and asymptotic estimates are available only under certain assumptions. 
This means that it is difficult to get a general estimate of the minimum number $R_c$ of canonical realizations required to distill a desired number $R_m$ of microcanonical graphs. 

Another criterion can be obtained by estimating the number $R_c$ of canonical realizations such that the microcanonical subset samples a desired \emph{fraction} $f_m$ (rather than a desired \emph{number} $R_m$) of all the $N_m(\mathbf{h}^*)$ microcanonical graphs. 
In this case, the knowledge of $N_m(\mathbf{h}^*)$ becomes unnecessary:
from the definition of $f_m$ we get
\begin{equation}
f_m\equiv\frac{R_m}{N_m(\mathbf{h}^*)}=
\frac{Q(\mathbf{h}^*)R_c}{N_m(\mathbf{h}^*)}=P(\mathbf{G}^*|\mathbf{\theta}^*)R_c
\label{eq:fm}
\end{equation}
The above formula shows that, if we want to sample a number $R_m$ of microcanonical realizations that span a fraction $f_m$ of the microcanonical ensemble, we need to sample a number 
\begin{equation}
R_c= \frac{f_m}{P(\mathbf{G}^*|\mathbf{\theta}^*)}
\label{eq:rc}
\end{equation} 
of canonical realizations and discard all the non-microcanonical ones.
This number can be extremely large, since $P(\mathbf{G}^*|\mathbf{\theta}^*)$ is very small, as we have already noticed. 
On the other hand, $f_m$ can be chosen to be very small as well.
To see this, let us for instance compare $f_m$ with the corresponding fraction 
\begin{equation}
f_c\equiv \frac{R_c}{N_c(\mathbf{h}^*)}
\end{equation}
of \emph{canonical} configurations sampled by $R_c$ realizations, where $N_c(\mathbf{h}^*)\gg N_m(\mathbf{h}^*)$ is the number of graphs in the canonical ensemble.
For all networks we considered in this paper, we showed that $R_c=1000$ realizations were enough to generate a good sample. 
This however corresponds to an extremely small value of $f_c$. For instance, for the binary interbank network we have  $f_c=1000/2^{N(N-1)/2}\approx 1.4\cdot 10^{-6920}$. 
We might therefore be tempted to choose the same small value also for $f_m$, and find the required number $R_c$ from eq. \eqref{eq:rc}.
However, the result is a value $R_c\ll 1$ (in the mentioned example, $R_c=2.8\cdot 10^{-3422}$), which clearly indicates that setting $f_m\equiv f_c$ (where $f_c$ is an acceptable canonical fraction) is inappropriate.
In general, $f_m$ should be much larger than $f_c$.

Importantly, we can show that, given a value  $R_c\gg 1$ that generates a good canonical sample, the subset of the $R_m$ microcanonical relations contained in the $R_c$ canonical ones spans a fraction $f_m$ of the microcanonical ensemble that is indeed much larger than $f_c$.
To see this, note that $P(\mathbf{G}^*|\mathbf{\theta}^*)$, being obtained with the introduction of the constraints $\mathbf{h}^*$, is necessarily much larger than the completely uniform probability $1/N_c(\mathbf{h}^*)$ over the canonical ensemble (corresponding to the absence of constraints).
This inequality implies that, if we compare $f_c$ with $f_m$ (both obtained with the same value of $R_c$), we find that
\begin{equation}
f_m=P(\mathbf{G}^*|\mathbf{\theta}^*)R_c\gg\frac{R_c}{N_c(\mathbf{h}^*)}=f_c
\label{eq:ff}
\end{equation}
The above expression shows that, even if only $R_m$ out of the (many more) $R_c$ canonical realizations belong to the microcanonical ensemble, the resulting microcanonical sampled fraction $f_m$ is still much larger than the corresponding canonical fraction $f_c$.
This non-obvious result implies that, in order to sample a microcanonical fraction that is much larger than the canonical fraction obtained with a given value of $R_c$, one does not need to increase the number of canonical realizations beyond $R_c$. 

The above considerations suggest that, under appropriate conditions, using our ``Max \& Sam'' method to sample the microcanonical ensemble might be competitive with the available microcanonical algorithms.
It should be noted that the value of  $R_c$ affects neither the preliminary search for the hidden variables $\mathbf{\theta}^*$, nor the calculation of the microcanonical averages over the $R_m$ final networks.
However, it does affect the number of checks  one has to make on the constraints to select the microcanonical networks.
The worst-case total number of checks is $O(R_c N)$, and performing such operation in a non-optimized way might slow down the algorithm considerably. 
A good strategy would be that of exploiting our analysis of the canonical fluctuations to identify the vertices for which it is more unlikely that the local constraint is matched exactly, and check these vertices first. This would allow one to identify, for each of the $R_c$ canonical realizations, the constraint-violating nodes at the earliest possible stage, and thus to abort the following checks for that particular network.
Implementing such an optimized microcanonical algorithm is however beyond the scope of this paper.

\section{Conclusions}
The definition and correct implementation of null models is a crucial issue in network analysis. 
When applied to real-world networks (that are generally strongly heterogeneous), the existing algorithms to enforce simple constraints on binary graphs become biased or time-consuming, and in any case difficult to extend to networks of different type (e.g. weighted or directed) and to more general constraints. 
We have proposed a fast and unbiased ``Max \& Sam'' method to sample several canonical ensembles of networks with various constraints. 

While canonical ensembles are believed to represent a mathematically tractable counterpart of microcanonical ones, they have not been used so far as a tool to sample networks with soft constraints, mainly because of the use of approximated expressions that result in ill-defined sampling probabilities.
Here, we have shown that it is indeed possible to use  exact expressions to correctly sample a number of canonical ensembles, from the standard case of binary graphs with given degree sequence to the more challenging models of directed and weighted graphs with given reciprocity structure or joint strength-degree sequence. Moreover, we have provided evidence that microcanonical and canonical ensembles of graphs with local constraints are not equivalent, and suggested that canonical ones can account for possible errors or missing entries in the data, while microcanonical ones do not.

Our algorithms are unbiased and efficient, as their computational complexity is $O(N^2)$ even for strongly heterogeneous networks. Canonical sampling algorithms may therefore represent an unbiased, fast, and more flexible alternative to their microcanonical counterparts. 
We have also illustrated the possibility to obtain an unbiased microcanonical method by discarding the realizations that do not match the constraints exactly. 
In our opinion, these findings might suggest new possibilities of exploitation of canonical ensembles as a solution to the problem of biased sampling in many other fields besides network science.

\appendix
\section*{Appendix: The ``Max \& Sam'' code}
An algorithm has been coded in various ways \cite{routine_address,routine_address2,routine_address3} in order to implement our sampling procedure for all the seven null models described in sec. \ref{sec:maxsam}. 
In what follows, we describe the Matlab implementation \cite{routine_address}.
A more detailed explanation accompanies the code in the form of a ``Read\_me'' file \cite{routine_address}. Here we briefly mention the main features.

The code can be implemented by typing a command having the typical form of a Matlab function, taking a number of different parameters as input.
The output of the algorithm is the numerical value of the \textit{hidden variables}, i.e. the vectors $\mathbf{x}$, $\mathbf{y}$ and $\mathbf{z}$ (where applicable) maximizing the likelihood of the desired null model (see sec. \ref{sec:maxsam}), plus a specifiable number of sampled matrices. 
The hidden variables alone allow the user to numerically compute the expected values of the adjacency matrix entries ($\langle a_{ij}\rangle\equiv p_{ij}$ and $\langle w_{ij}\rangle$), as well as the expected value of the constraints (as a check of its consistency with the observed value),  according to the specific definition of each model. 
Moreover, the user can obtain as output any number of matrices (networks) sampled from the desired ensemble.
These matrices are sampled in an unbiased way from the canonical ensemble corresponding to the chosen null model, using the relevant random variables as described in sec. \ref{sec:maxsam}.

The command to be typed is the following (more details can be found in the ``Read\_me'' file \cite{routine_address}):

\begin{framed}
\parbox{\textwidth}{\tt output = MAXandSAM(method, Matrix, Par,}
\parbox{\textwidth}{\tt \hspace{3.7cm} List, eps, sam, x0new)}
\end{framed}

The first parameter (\texttt{method}) can be entered by typing the acronym associated with the selected null model: 
\begin{itemize}
\item \textbf{UBCM} for the Undirected Binary Configuration Model, preserving the degree sequence ($\{k_i\}_{i=1}^N$) of an undirected binary network $\mathbf{A}^*$ (see sec. \ref{sec:UBCM}); 

\item\textbf{DBCM} for the Directed Binary Configuration Model, preserving the in- and out-degree sequences ($\{k_i^{in}\}_{i=1}^N$ and $\{k^{out}_i\}_{i=1}^N$) of a directed binary network $\mathbf{A}^*$ (see sec. \ref{sec:DBCM}); 

\item \textbf{RBCM} for the Reciprocal Binary Configuration Model, preserving the reciprocated, incoming non-reciprocated and outgoing non-reciprocated degree sequences ($\{k_i^{\leftrightarrow}\}_{i=1}^N$, $\{k_i^{\leftarrow}\}_{i=1}^N$ and $\{k^{\rightarrow}_i\}_{i=1}^N$) of a directed binary network $\mathbf{A}^*$ (see sec. \ref{sec:RBCM}); 

\item \textbf{UWCM} for the Undirected Weighted Configuration Model, preserving the strength sequence ($\{s_i\}_{i=1}^N$) of an undirected weighted network $\mathbf{W}^*$ (see sec. \ref{sec:UWCM});
 
\item \textbf{DWCM} for the Directed Weighted Configuration Model, preserving the in- and out-strength sequences ($\{s_i^{in}\}_{i=1}^N$ and $\{s^{out}_i\}_{i=1}^N$) of a directed weighted network $\mathbf{W}^*$ (see sec. \ref{sec:DWCM}); 

\item \textbf{RWCM} for the Reciprocal Weighted Configuration Model, preserving the the reciprocated, incoming non-reciprocated and outgoing non-reciprocated strength sequences ($\{s_i^{\leftrightarrow}\}_{i=1}^N$, $\{s_i^{\leftarrow}\}_{i=1}^N$ and $\{s^{\rightarrow}_i\}_{i=1}^N$) of a directed weighted network $\mathbf{W}^*$ (see sec. \ref{sec:RWCM});

\item \textbf{UECM} for the Undirected Enhnaced Configuration Model, preserving both the degree and strength sequences ($\{k_i\}_{i=1}^N$ and $\{s_i\}_{i=1}^N$) of an undirected weighted network $\mathbf{W}^*$ (see sec. \ref{sec:UECM}).
\end{itemize}

The second, third and fourth parameters (\texttt{Matrix}, \texttt{Par} and \texttt{List} respectively) specify the format of the input data (i.e. of $\mathbf{A}^*$ or $\mathbf{W}^*$). Different data formats can be taken as input: 
\begin{itemize}
\item \texttt{Matrix} for a (binary or weighted) matrix representation of the data, i.e. if the entire adjacency matrix is available;
\item \texttt{List} for an edge-list representation of the data, i.e. a $L\times 3$ matrix ($L$ being the number of links) with the first column listing the starting node, the second column listing the ending node and the third column listing the weight (if available) of the corresponding link; 
\item \texttt{Par} when only the constraints' sequences (degrees, strengths, etc.) are available. 
\end{itemize}
In any case, the two options that are not selected are left empty, i.e. their value should be ``\textbf{[ ]}''.
We stress that the likelihood maximization procedure (or the solution of the corresponding system of equations making the gradient of the likelihood vanish), which is the core of the algorithm, only needs the observed values of the chosen constraints to be implemented. However, since different representations of the system are available, we have chosen to exploit them all and to let the user choose the most appropriate to the specific case.
For instance, in network reconstruction problems \cite{myenhanced} one generally has empirical access only to the local properties (degree and/or strength) of each node, and the full adjacency matrix is unknown.

The fifth parameter (\texttt{eps}) controls for the maximum allowed relative error between the observed and the expected value of the constraints. According to this parameter, the code solves the entropy-maximization problem by either just maximizing the likelihood function or also improving this first outcome solution by further solving the associated system. Even if this choice might strongly depend on the observed data, the value $\epsilon=10^{-6}$ works satisfactorily in most cases.

The sixth parameter (\texttt{sam}) is a boolean variable allowing the user to extract the desired number of matrices from the chosen ensemble (using the probabilities $p_{ij}$). The value ``0'' corresponds to no sampling: with this choice, the code gives only the hidden variables as output. 
If the user enters ``1'' as input value, the algorithm will ask him/her to enter the number of desired matrices (after the hidden variables have been found). 
In this case, the code outputs both the hidden variables and the sampled matrices, the latter in a \texttt{.mat} file called \texttt{Sampling.mat}.

The seventh parameter (\texttt{x0new}) is optional and has been introduced to further refine the solution of the UECM \cite{myenhanced} in the very specific case of networks having, at the same time, big outliers in the strength distribution and a narrow degree distribution. In this case, the optional argument \texttt{x0new} can be inputed with the previously obtained output: in so doing, the code will solve the system again, by using the previous solution as initial point. This procedure can be iterated until the desired precision is reached. Note that, since \texttt{x0new} is an \emph{optional} parameter, it is not required to enter ``\textbf{[ ]}'' when the user does not need it (differently e.g. from the data format case).

\begin{acknowledgements}
DG acknowledges support from the Dutch Econophysics Foundation (Stichting Econophysics, Leiden, the Netherlands) with funds from beneficiaries of Duyfken Trading Knowledge BV, Amsterdam, the Netherlands. 
This work was also supported by the EU project MULTIPLEX (contract 317532) and the Netherlands Organization for Scientific Research (NWO/OCW).
\end{acknowledgements}

\end{document}